\documentclass[notitlepage,prd,preprintnumbers,longbibliography,showpacs]{revtex4-1}
\usepackage[%
colorlinks=true,
urlcolor=blue,
linkcolor=blue,
citecolor=blue
]{hyperref}
\usepackage{slashed,color,amsmath,amssymb,mathrsfs}
\usepackage{graphicx}
\usepackage{hyperref}
\usepackage{longtable}
\usepackage{array}
\usepackage{accents}
\usepackage{tensor}
\usepackage{longtable}
\usepackage{booktabs}
\usepackage{multirow}
\usepackage{bm}
\usepackage{subfigure}
\graphicspath{{figures/}}

\newcommand{\Mev}{\mathrm{MeV}}
\allowdisplaybreaks

\begin{document}
	
\title[]{New method for fitting the low-energy constants in chiral perturbation theory}
\author{Qin-He Yang$^{1}$}
\email{yqh@st.gxu.edu.cn}
\author{Wei Guo$^{1}$}
\author{Feng-Jun Ge$^{2}$}
\author{Bo Huang$^{1}$}
\author{Hao Liu$^{1}$}
\author{Shao-Zhou Jiang$^{1}$}
\email[Corresponding author.]{jsz@gxu.edu.cn}
\affiliation{$^{1}$ Key Laboratory for Relativistic Astrophysics, Department of Physics,Guangxi University,	Nanning 530004, People's Republic of China}
\affiliation{$^{2}$ Institute of Applied Physics and Computational Mathematics, Beijing 100094, People's Republic of China}

\begin{abstract}
	
A new set of the next-to-leading order (NLO) and the next-to-next-to-leading order (NNLO) low-energy constants $L_i^r$ and $C_i^r$ in chiral perturbation theory is obtained. These values are computed using the new experimental data with a new calculation method. This method combines the traditional global fit and Monte Carlo method together. The higher order contributions are estimated with this method. The theoretical values of the observables provide good convergence at each chiral dimension, except for the NNLO values of the $\pi K$ scattering lengths $a_0^{3/2}$ and $a_0^{1/2}$. The fitted values for $L_i^r$ at NLO are close to their results with the new method at NNLO; i.e., these $L_i^r$ are nearly order-independent in this method. The estimated ranges for $C_i^r$ are consistent with those in the literature, and their possible upper or/and lower boundaries are given. The values of some linear combinations of $C_i^r$ are also given, and they are more reliable. If one knows a more exact value $C_i^r$, another $C_i^r$ can be obtained by these values.
\end{abstract}

\maketitle
\section{Introduction}\label{intro}

Chiral perturbation theory (ChPT) is an important tool to study the low-energy pseudoscalar mesonic interactions. The main idea of ChPT comes from the fact that QCD possesses an $SU(3)_L\times SU(3)_R$ flavor symmetry in the chiral limit in which the light quark are considered massless. This symmetry is spontaneously broken into the subgroup $SU(3)_V$ and eight massless pseudoscalar Goldstone bosons arise. These pseudoscalar Goldstone bosons are considered to be the lightest eight pseudoscalar mesons ($\pi$, $K$ and $\eta$). The small masses of these pseudoscalar mesons come from the small light-quark masses. On the other hand, the fundamental interaction between the pseudoscalar mesons in ChPT can be considered as an effective interaction in the low energy, i.e. ChPT is only an effective theory. The only restriction is symmetry, such as chiral symmetry, parity symmetry, charge conjugation symmetry, and so on. There exist an infinite number of terms satisfying these symmetries, and an infinite number of unknown parameters called low-energy constants (LECs) that correspond to these terms. The details of the strong interaction are hidden in these LECs. The Weinberg power-counting scheme organizes the most important terms to be considered first, the second important ones to be calculated secondly, and so on \cite{Weinberg:1978kz}. Generally, physical quantities in ChPT are calculated order by order (chiral dimension). One order provides about a $p/\Lambda_{\chi}$ factor, where $p$ is the typical scale of the momentum and $\Lambda_{\chi}\sim 1$ GeV is the scale related to chiral symmetry breaking. For the three-flavor ChPT, there exist 2, 10+2, 90+4, and 1233+21 LECs in the $\mathcal{O}(p^2)$, $\mathcal{O}(p^4)$, $\mathcal{O}(p^6)$, and $\mathcal{O}(p^8)$ order, respectively \cite{Gasser:1984gg,Gasser:1983yg,Bijnens:1999sh,Bijnens:2018lez}. The numbers after the plus signs are related to the contact terms. It shows that the numbers increase rapidly with the growth of the chiral dimension. Nevertheless, ChPT can not determine these LECs by itself. Without LECs, most physical quantities would not be calculated numerically and ChPT would lose most of its predictions. Hence, numerical values of LECs play an important role in ChPT. There are a lot of methods to obtain LECs, such as global fit \cite{Bijnens:1994ie,Amoros:2000mc,Bijnens:2011tb,Bijnens:2014lea}, lattice QCD \cite{Dowdall:2013rya,Bazavov:2010hj,Bazavov:2009fk,Bazavov:2009bb,Bernard:2009ds}, chiral quark model \cite{Jiang:2009uf,Jiang:2015dba}, resonance chiral theory \cite{Amoros:1999dp,Knecht:2001xc}, sum rules \cite{Golterman:2014nua}, holographic QCD \cite{Colangelo:2012ipa}, dispersion relations \cite{Guo:2007ff,Guo:2007hm,Guo:2009hi}, and so on. Each method has its advantage and application domain. However, at present, most of them only obtain a part of LECs at/to a given order and the higher-order contributions are neglected. Most numerical results satisfy the power-counting scheme, but there also exist some exceptions (see the discussion below). However, the calculation at/to a given order sometimes may not give a very good prediction. It leads to numerical values of some LECs that may have large errors. This is one possibility why some LECs have large errors in some references. One motivation for this paper is to obtain some LECs with the higher-order contributions in order to narrow the errors of some NLO LECs.

In this paper, some three-flavor LECs will be obtained with a new method, which is similar to the traditional global fit method but with some improvements. The traditional global fit method seems simpler than the pure calculation by the background theory. This method is also much closer to the experiment, because it fits the experimental data directly. The theoretical values and errors can be obtained simultaneously without the background theory or any other physical model. Usually, $\chi^2$ in the fit is as small as possible, and the corresponding LECs are the result. However, the global fit method needs sufficient theoretical calculations in ChPT, some of them which may be in the high order with tedious loop-diagram calculations. Its precision is limited by the number and the accuracy of experimental data. So far, a lot of research arises and some LECs have been fitted. $L_1^r$, $L_2^r$ and $L_3^r$ are obtained by fitting $K_{\ell 4}$ form factors and $\pi \pi$ scattering lengths \cite{Bijnens:1994ie}. $L_i^r(i=1,2,3,5,7,8)$ are obtained by fitting the quark-mass ratio $m_s/\hat m$, the decay constant ratio $F_K/F_\pi$ and $K_{\ell4}$ form factors \cite{Amoros:2000mc}. About ten years later, Ref. \cite{Bijnens:2011tb} added $\pi \pi$ scattering lengths ($a_0^0$ and $a_0^2$), $\pi K$ scattering lengths ($a_0^{1/2}$ and $a_0^{1/3}$) and the threshold parameters of the scalar form factor ($\langle r^2\rangle_S^\pi$ and $c_S^\pi$) in the fit. Recently, Ref. \cite{Bijnens:2014lea} added two-flavor LECs in the fit. The last two references not only obtain the next-to-leading order [NLO, $\mathcal{O}(p^4)$] LECs $L_i^r$ but also estimate a part of LECs $C_i^r$ in the next-to-next-to-leading order [NNLO, $\mathcal{O}(p^6)$]. Nevertheless, their results only make use of the theoretical expansion to a finite order (NNLO). The higher-order contributions are ignored. There also exist some other problems in Ref. \cite{Bijnens:2014lea}:
\begin{enumerate}
\item Some NLO fitted values of $L_i^r$ are quite different from those obtained at NNLO. For example, $L_1^r=0.53(06)\times 10^{-3}$ at a NNLO fit, which is about half of its NLO fitted value $L_1^r=1.0(1)\times 10^{-3}$. Nevertheless, $L_1^r$ is a constant. Its true value is order independent. Its fitted value should not depend on the order as far as possible.

\item The higher-order effects are not taken into account. If the higher-order corrections are considered, some physical quantities may change largely, such as $a_0^{1/2}$ and $a_0^{3/2}$. Reference \cite{Bijnens:2014lea} tells us that the numerical values of $a_0^{1/2}$ and $a_0^{3/2}$ at NNLO are larger than the NLO results. Hence, the higher-order effects should have a big impact on the low-order LECs. In addition, one can determine whether a set of LECs have reliable values, if the higher-order effects are known. If the possible higher-order effects are smaller than the low-order ones, the set of LECs should be considered more reliable. Hence, the higher-order effects should play an important role in the fit.

\item The fitted $\chi^2/\mathrm{d.o.f.}$ is approximately equal to $1.0/10$ at NNLO fit. There seems to be an overfitting problem. A larger $\chi^2$ could give a wider range of $C_i^r$. Some $C_i^r$ in this wider range may solve the two problems discussed above.

\item The $\pi K$ scattering lengths $a_0^{1/2}$ and $a_0^{3/2}$ have a poor convergence. Compared with the NLO results, their NNLO values are too large.

\item Some original data about $C_i^r$ \cite{Bijnens2019} are very close to the final results \cite{Bijnens:2014lea}. The differences are less than $10^{-12}$. We guess these LECs might be dependent on the boundaries.
\end{enumerate}

In this paper, we attempt to solve the first three problems and try to find out why the other two problems arise. With some reasonable hypotheses, a new method for fitting LECs is introduced and a set of $L_i^r$ and $C_i^r$ are obtained. Because the number of constraint conditions is not large enough, only the ranges of $C_i^r$ are obtained, but the values of $L_i^r$ are more precise.

This paper is organized as follows: In Sec. \ref{Sec:III}, some hypothesises are introduced and the following calculations are based on these hypothesises. In Sec. \ref{input}, all the experimental data involved are given. Section \ref{m1} introduces a modified global fit method with the higher-order estimation. In Sec. \ref{r1}, some rough values of $L_i^r$ are given and the convergences of some observables are also presented. In Sec. \ref{m2}, a new method is introduced, which can compute more reasonable values of $L_i^r$ and estimate the values of $C_i^r$. Section \ref{r2} gives the results of $L_i^r$ and $C_i^r$ with this new method. A short summary is given in Sec. \ref{sum}

\section{The low-energy constants and their hypotheses}\label{Sec:III}
In ChPT, without the contact terms, for the three-flavor case, there are 10 LECs $L_i$ at NLO and 90 LECs $C_i$ at NNLO; for the two-flavor case, 7 LECs $l_i$ exist at NLO and 52 LECs $c_i$ exist at NNLO. Their renormalized values $L_i^r$, $C_i^r$, $l_i^r$ and $c_i^r$ are defined in Refs. \cite{Gasser:1983yg,Gasser:1984gg,Scherer:2012xha,Bijnens:2013zc,Bijnens:2014lea,Bijnens:1999hw}. Some scale independent $\bar{l}_i$ \cite{Gasser:1983yg} are used frequently. This paper will determine the values of 8 $L_i^r\;(i=1,\ldots,8)$ and 38 $C_i^r$ (the values of $i$ can be found in Table \ref{table5}) at the renormalization scale $\mu=0.77$GeV. Four $\bar{l}_i\;(i=1,2,3,4)$ will be used in the estimation as observables, and none of the $c_i$ will be used. The following notations are the same as those in Ref. \cite{Bijnens:2014lea}.

Due to the experimental condition and the theoretical calculation, the experimental data and the relevant NNLO analytical results are lacking. Until now, only 17 observables have been used in the NNLO global fit \cite{Bijnens:2014lea}. Theoretically, it is impossible to obtain all LECs very accurately with only these 17 observables. Hence, we have different requirements for these NLO and NNLO LECs. For $L_i^r$, their values need to be as precise as possible, because their number (8) is much less than 17. On the other hand, although 17 is less than 38, it does not seem too small to estimate $C_i^r$. With the help of some reasonable hypotheses, the intervals of $C_i^r$ can be limited to some reliable ranges at least. To achieve these goals, the following hypotheses are introduced to limit the feasible ranges of the LECs:
\begin{enumerate}
  \item \label{hy1} Chiral expansion for most observables is assumed to have good convergence. Observables are expanded by the momentum and the quark masses in ChPT. Any observable is calculated (chiral) order by order. The high-order value should be small enough compared with the low-order one. This is a theoretical assumption in the effective theory. For most observables, the LO values give the greatest contributions. The NLO and the NNLO values are smaller and smaller. The sum of the unknown higher-order contributions, which is also called truncation error, should be smaller than the NNLO values. There should exist some exceptions. These exceptions will be considered separately.
  
	\item \label{hy2} All $L_i^r$ are assumed to be stable. In other words, the values of $L_i^r$ obtained at both NLO and NNLO should be almost unchanged. This is because that all LECs are constants, and they are independent of the different computational methods. According to hypothesis \ref{hy1}, the contributions at NNLO and the truncation error would be small enough. These small contributions only lead to a small variation of $L_i^r$. However, this does not always work. In Ref. \cite{Bijnens:2014lea}, some NNLO fitted $L_i^r$ have large differences from the NLO fitted ones. For example, the NNLO fitted value of $L_1^r=0.53(04)\times10^{-3}$ is half of its NLO fitted value $1.0(1)\times10^{-3}$. The deviation from the NLO value is about $5\sigma$. The reason is that these values are only fitted at a given order and the truncation errors are neglected. For some $L_i^r$, this effect is not very obvious, but for the other ones, it may have a large impact on their values. This hypothesis will be used for constraining the ranges of $L_i^r$ at NNLO calculation, which are assumed to be close to the NLO fitted values. We choose the difference between the NLO and the NNLO fitted values to be less than 20\% in this paper.

	\item \label{hy3} All $C_i^r$ should be consistent with those obtained from the other references; i.e., their values can not deviate too much from those in the other references. Because these $C_i^r$ are derived by different approximations, their results should be close to the real values. We consider all of their results reasonable. Hence, we have no reason to deny any of their results. These $C_i^r$ are treated as boundary conditions to constrain the ranges of $C_i^r$. Compared with $L_i^r$, the number of $C_i^r$ is very large and $C_i^r$ are hard to be determined. Their values appear less than $L_i^r$ in the literature. Appendix \ref{app:C} presents all relevant $C_i^r$ that we can find. The distributions of most $C_i^r$ are wide. The true values of $C_i^r$ are assumed to be in or close to these wide ranges.
\end{enumerate}

\section{Observables, inputs and $\chi^2$}\label{input}
This paper is based on Refs. \cite{Bijnens:2011tb,Bijnens:2014lea}, which adopt a global fit method to obtain $L_i^r$ and use a random walk algorithm to estimate $C_i^r$. For the NLO fit, the following 12 observables are used. The mass ratio $m_s/\hat m$ can be calculated according to pion and kaon masses ($m_s/\hat m|_1$) or pion and eta masses ($m_s/\hat m|_2$) \cite{Amoros:1999dp,Amoros:2000mc,Bijnens:2011tb,Bijnens2019}, where $m_s$ is the strange quark mass and $\hat m=(m_u+m_d)/2$ is the isospin doublet quark mass. The ratio of the kaon decay constant $F_K$ to the pion decay constant $F_\pi$ ($F_K/F_\pi$) is also used in the fit \cite{Amoros:1999dp,Bijnens:2011tb,Bijnens:2014lea,Bijnens2019}, which eliminates the unknown constant $F_0$. There exist two form factors $F$ and $G$ in the $K_{\ell 4}$ decay; their values and slops at threshold ($f_s$, $g_p$, $f'_s$ and $g'_s$) \cite{Amoros:2000mc} are also considered in the fit. The $\pi\pi$ scattering lengths $a_0^0$ and $a_0^2$ \cite{Bijnens:2004eu,Bijnens2019}, the $\pi K$ scattering lengths $a_0^{1/2}$ and $a_0^{3/2}$, and the pion scalar radius $\langle r^2\rangle_S^\pi$ in the form factor $F_S^\pi(t)$ \cite{Bijnens:2003xg} are also included. With these 12 observables, eight $L_i^r (i=1,\ldots, 8)$ will be fitted. The other five observables are added at the NNLO fit, they are the pion scalar curvature $c_S^\pi$ \cite{Bijnens:2003xg} and four two-flavor LECs $\bar l_i\,(i=1,\ldots,4)$ \cite{Gasser:2007sg}. In this paper, we also adopt the same observables in the fit and only update some experimental or theoretical data. All the calculations are in three flavor ChPT. The analytical results can be found in the above references. In the calculation, these observables are treated as independent ones. The calculations are related to 8 $L_i^r$ and 38 $C_i^r$. The total number 46 is larger than the number of observables. We will use a different method to obtain them. The renormalization scale $\mu$ is chosen to be 0.77 GeV in this paper.

The values of the meson masses and the pion decay constant are
\begin{align}
\nonumber
m_\pi^{\pm}=&139.57061(24)\,\Mev,\quad m_\pi^0=134.9770(5)\,\Mev,\quad m_\eta=547.862(17)\,\Mev,\\
m_K^{\pm}=&493.677(16)\,\Mev,\quad m_K^0=497.611(13)\,\Mev,\quad F_\pi=92.3\pm 0.1\,\Mev.\label{equ:12}
\end{align}
The average kaon mass is
\begin{align}
m_{K\,\mathrm{av}}=494.50\,\Mev.\label{equ:13}
\end{align}
which is used in the calculation for the pion and kaon decay constants and the pseudoscalar meson masses \cite{Bijnens:1996kk}.

The values of $m_s/\hat m$ and $F_K/F_\pi$ are \cite{Tanabashi:2018oca}
\begin{align}
\frac{m_s}{\hat m}=27.3_{-1.3}^{+0.7},\quad\frac{F_K}{F_\pi}=1.199\pm 0.003.\label{equ:14}
\end{align}

For $K_{\ell4}$ form factors $F$ and $G$, their slope and value at threshold are \cite{Tanabashi:2018oca}
\begin{align}
\nonumber
f_s=5.712\pm 0.032,\quad f_s^{\prime}=0.868\pm 0.049,\\
g_p=4.958\pm 0.085,\quad g^{\prime}_p=0.508\pm 0.122.\label{equ:15}
\end{align}

The latest results for $\pi\,\pi$ scattering lengths are given in Ref. \cite{Batley:2010zza}, which are based on the analysis of $K_{e4}$ data. Their values are
\begin{align}
a_0^{0}=0.2196\pm 0.0034,\quad a_0^{2}=-0.0444\pm 0.0012.\label{equ:16}
\end{align}

For $\pi\, K$ scattering lengths, Ref. \cite{Adeva:2017oco} gives the most recent experimental value for the S-wave isospin-odd $\pi K$ scattering length $a_0^-=|a_0^{1/2}-a_0^{3/2}|/3$ , but we have not found any update of $a_0^{1/2}$ or $a_0^{3/2}$ separately.
Hence, we still use the same data as those in Refs. \cite{Buettiker:2003pp,Bijnens:2014lea},
\begin{align}
a_0^{1/2}m_\pi=0.224\pm 0.022,\quad a_0^{3/2}m_\pi=-0.0448\pm 0.0077.\label{equ:17}
\end{align}

Since no update has been found, the scalar radius $\langle r^2\rangle_S^\pi$ and the scaler curvature $c_S^\pi$ of the pion scalar form factor are the same as those in Ref. \cite{Bijnens:2014lea}. Their values are based on the dispersion analysis \cite{Donoghue:1990xh,Moussallam:1999aq},
\begin{align}
\langle r^2\rangle_S^\pi=0.61\pm 0.04\,\mathrm{fm}^2,\quad c_S^\pi=11\pm 1\,\mathrm{GeV}^{-4}.\label{equ:18}
\end{align}

For two-flavor LECs $\bar l_i\,(i=1,\ldots,4)$, the values of $\bar l_1$ and $\bar l_2$ are chosen \cite{Colangelo:2001df},
\begin{align}
\bar l_1=-0.4\pm 0.6,\quad \bar l_2=4.3\pm 0.1,\label{equ:19}
\end{align}
which are the same as those in Ref. \cite{Bijnens:2014lea}. For $\bar l_3$ and $\bar l_4$, Ref. \cite{Bijnens:2014lea} uses the average of lattice results \cite{Aoki:2013ldr,Durr:2013goa} and the continuum results \cite{Colangelo:2001df,Gasser:1983yg}. At this time, the lattice results in Ref. \cite{Durr:2013goa} are not included in Flavour Lattice Averaging Group (FLAG) average \cite{Aoki:2013ldr}. The most recent FLAG data \cite{Aoki:2019cca} provide the following averages
\begin{align}
\nonumber
\bar l_3|_{N_f=2}=3.41(82),\quad \bar l_3|_{N_f=2+1}=3.07(64),\quad \bar l_3|_{N_f=2+1+1}=3.53(26),\\
\bar l_4|_{N_f=2}=4.40(28),\quad \bar l_4|_{N_f=2+1}=4.02(45),\quad \bar l_4|_{N_f=2+1+1}=4.73(10).\label{equ:20}
\end{align}
The values in Eq. \eqref{equ:20} have included the results in Ref. \cite{Durr:2013goa}. A new estimate according to Eq. \eqref{equ:20} and Refs. \cite{Colangelo:2001df,Gasser:1983yg} is
\begin{align}
\bar l_3=3.2\pm 0.7,\quad \bar l_4=4.4\pm 0.2.\label{equ:21}
\end{align}
The $\bar{l}_i$ values in Eqs. \eqref{equ:19} and \eqref{equ:21} are adopted in our fit.

Equations. \eqref{equ:12} -- \eqref{equ:19} and \eqref{equ:21} are all physical quantities used in our calculation.

The objective function in the estimation, $\chi^2$, is the same as those in Refs. \cite{Amoros:2000mc,Bijnens:2011tb,Bijnens:2014lea},
\begin{align}
\chi^2=\sum_i\chi_i^2=\sum_i\left(\frac{X_{i({\rm th})}-X_{i({\rm exp})}}{\Delta X_i}\right)^2,\label{chi2}
\end{align}
where $X_{i({\rm exp})}$ are the experimental values, $X_{i({\rm th})}$ are the theoretical estimates and $\Delta X_i$ are the experimental errors. Generally, $\chi^2$ is as small as possible. This function is a criterion to judge whether the LECs are reasonable or not. The errors of $L_i^r$ give $\Delta\chi^2=1$, assuming the quadratic approximation is near the minimum.

The above $\chi^2$ contains both $\bar{l}_{i}$ and the $\pi\,\pi$ scattering lengths $a_0^{0}$ and $ a_0^{2}$. However, in the two-flavor ChPT, $a_0^{0}$ and $ a_0^{2}$ can be calculated by $\bar{l}_i$ to NLO. The higher-order contributions are small. Hence, $\bar{l}_i$ are statistically correlated to $a_0^{0}$ and $ a_0^{2}$. Strictly speaking, the inverse covariance matrix of $\bar{l}_i$ needs to be considered in Eq. \eqref{chi2}, but it is hard to be determined. We have not found it in the literature. Hence, the covariance matrix is ignored in Eq. \eqref{chi2}, and Eq. \eqref{chi2} does not lead to a normal $\chi^2$ distribution. Equation \eqref{chi2} is only a modified $\chi^2$ fitting. The contributions from $\bar{l}_i$ should be regarded as extra constraints of $L_i^r$ and $C_i^r$ in $\chi^2$ but not independent influence. The contributions from $\bar{l}_i$ are about $15\%$ of the total $\chi^2$. Their influence is not very large and has little impact on the final results.

\section{Method I: A modified global fit for obtaining $L_i^r$}\label{m1}
In this section, a modified global fit method is introduced, which contained the higher-order estimates in the fit. This method is only for estimating the values of $L_i^r$. The calculating process is similar to that in Refs. \cite{Bijnens:2011tb,Bijnens:2014lea}. Only the differences are explained. 

\subsection{Chiral expansions}

In ChPT, physical quantities are calculated order by order, but some quantities described above are mixed by different orders. In order to pick out the exact contributions from different orders, they need to be expanded order by order. 

The expansion for the ratio $F_K/F_\pi$ to the NNLO is \cite{Bijnens:2014lea}
\begin{align}
\frac{F_K}{F_\pi}\approx&\underbrace{1}_{\text{LO}}+\underbrace{\Big(\frac{F_K}{F_0}\Big)_4-\Big(\frac{F_\pi}{F_0}\Big)_4}_{\text{NLO}}+\underbrace{\Big(\frac{F_K}{F_0}\Big)_6-\Big(\frac{F_\pi}{F_0}\Big)_6-\Big(\frac{F_\pi}{F_0}\Big)_4\Big[\Big(\frac{F_K}{F_0}\Big)_4-\Big(\frac{F_\pi}{F_0}\Big)_4\Big]}_{\text{NNLO}}.\label{equ:22}
\end{align}
Hereafter, the subscript 2, 4, 6 and 8 are represented the contribution at LO, NLO, NNLO and NNNLO, respectively.

The quark-mass ratio $m_s/\hat m$ can be calculated according to the LO pion and kaon masses or the LO pion and eta masses
\begin{align}
\frac{m_s}{\hat m}\Big|_1=\frac{2m_{K2}^2-m_{\pi2}^2}{m_{\pi2}^2},\quad \frac{m_s}{\hat m}\Big|_2=\frac{3m_{\eta2}^2-m_{\pi2}^2}{2m_{\pi2}^2}.\label{equ:23}
\end{align}
Their expansions are
\begin{align}
\nonumber
\frac{m_s}{\hat m}\Big|_1\approx&\frac{2[m_K^2-(m_K^2)_4-(m_K^2)_6]-[m_\pi^2-(m_\pi^2)_4-(m_\pi^2)_6]}{[m_\pi^2-(m_\pi^2)_4-(m_\pi^2)_6]}\\
\approx&\underbrace{\frac{2m_K^2-m_\pi^2}{m_\pi^2}}_{\text{LO}}+\underbrace{\frac{2m_K^2(m_\pi^2)_4}{m_\pi^4}-\frac{2(m_K^2)_4}{m_\pi^2}}_{\text{NLO}}+\underbrace{\frac{2m_K^2(m_\pi^2)_4^2}{m_\pi^6}-\frac{2(m_K^2)_4(m_\pi^2)_4}{m_\pi^4}+\frac{2m_K^2(m_\pi^2)_{6}}{m_\pi^4}-\frac{2(m_K^2)_{6}}{m_\pi^2}}_{\text{NNLO}},\label{equ:24}\\\nonumber
\frac{m_s}{\hat m}\Big|_{2}\approx&\frac{3[m_\eta^2-(m_\eta^2)_4-(m_\eta^2)_6]-[m_\pi^2-(m_\pi^2)_4-(m_\pi^2)_6]}{2[m_\pi^2-(m_\pi^2)_4-(m_\pi^2)_6]}\\
\approx&\underbrace{\frac{3m_\eta^2-m_\pi^2}{2m_\pi^2}}_{\text{LO}}+\underbrace{\frac{3m_\eta^2(m_\pi^2)_4}{2m_\pi^4}-\frac{3(m_\eta^2)_4}{2m_\pi^2}}_{\text{NLO}}+\underbrace{\frac{3m_\eta^2(m_\pi^2)_4^2}{2m_\pi^6}-\frac{3(m_\eta^2)_4(m_\pi^2)_4}{2m_\pi^4}+\frac{3m_\eta^2(m_\pi^2)_{6}}{2m_\pi^4}-\frac{3(m_\eta^2)_{6}}{2m_\pi^2}}_{\text{NNLO}}.\label{equ:25}
\end{align}

$\langle r^2\rangle_S^\pi$ and $c_S^\pi$ are related to the differential of form factors $F_S^\pi(t)$. Their expansions are
\begin{align}
\nonumber
\langle r^2\rangle_S^\pi=&\frac{6}{F_S^\pi(0)}\frac{d}{dt}F_S^\pi(t)|_{t=0}\\\nonumber
\approx&\underbrace{0}_{\text{LO}}+\underbrace{6\Big(\frac{F_S^\pi}{2B_0}\Big)'_4}_{\text{NLO}}+\underbrace{6\Big[\Big(\frac{F_S^\pi}{2B_0}\Big)'_{6}-\Big(\frac{F_S^\pi}{2B_0}\Big)'_4\Big(\frac{F_S^\pi}{2B_0}\Big)_4\Big]}_{\text{NNLO}}\\
&+\underbrace{6\Big\{\Big(\frac{F_S^\pi}{2B_0}\Big)'_8-\Big(\frac{F_S^\pi}{2B_0}\Big)'_4\Big[\Big(\frac{F_S^\pi}{2B_0}\Big)_{6}+\Big(\frac{F_S^\pi}{2B_0}\Big)_4^2\Big]-\Big(\frac{F_S^\pi}{2B_0}\Big)_4\Big(\frac{F_S^\pi}{2B_0}\Big)'_{6}\Big\}}_{\text{NNNLO}}\bigg|_{t=0},\label{equ:26}\\
c_S^\pi=&\frac{1}{2}\frac{1}{F_S^\pi(0)}\frac{d^2}{dt^2}F_S^\pi(t)|_{t=0}\nonumber\\
\approx&\underbrace{0}_{\text{LO}}+\underbrace{\frac{1}{2}\Big(\frac{F_S^\pi}{2B_0}\Big)''_4}_{\text{NLO}}+\underbrace{\frac{1}{2}\Big[\Big(\frac{F_S^\pi}{2B_0}\Big)''_{6}-\Big(\frac{F_S^\pi}{2B_0}\Big)''_4\Big(\frac{F_S^\pi}{2B_0}\Big)_4\Big]}_{\text{NNLO}}\nonumber\\
&+\underbrace{\frac{1}{2}\Big\{\Big(\frac{F_S^\pi}{2B_0}\Big)''_8-\Big(\frac{F_S^\pi}{2B_0}\Big)''_4\Big[\Big(\frac{F_S^\pi}{2B_0}\Big)_{6}+\Big(\frac{F_S^\pi}{2B_0}\Big)_4^2\Big]-\Big(\frac{F_S^\pi}{2B_0}\Big)_4\Big(\frac{F_S^\pi}{2B_0}\Big)''_{6}\Big\}}_{\text{NNNLO}}\bigg|_{t=0}.\label{equ:27}
\end{align}
The first terms in Eqs. \eqref{equ:26} and \eqref{equ:27} are both equal to zero; it is due to the fact that the scalar form factor at LO is independent on $t$.

\subsection{The estimation at the higher order}\label{Sec:IV}

In the previous fitting methods \cite{Amoros:2000mc,Bijnens:2011tb,Bijnens:2014lea}, the influences from the higher orders have not been taken into account. Although the truncation errors should be very small according to hypothesis \ref{hy1}, it is also worth to evaluate the influences from the higher orders according to hypothesis \ref{hy2}. Higher-order contributions may have a big impact on some values of $L_i^r$. However, the contributions of the order beyond NNLO are absolutely unknown, and they need to be estimated in other ways. Reference \cite{Furnstahl:2015rha,Melendez:2017phj} provide a method for the quantitative estimation of the truncation errors, which is based on Bayesian method. They assume that the expansion coefficients of the observables in the effective field theory are of natural size, and their distributions are symmetric about the origin. The distribution of the truncation errors is also symmetric about the origin. The confidence intervals can be obtained in several ways. This assumption leads to a zero center value and a nonzero uncertainty band. In practice, contributions from the higher orders may not be equal to zero. Some nonzero estimates need to be obtained, but we do not find an effective way to use the nonzero uncertainty band by the computer. In this section, a method for estimating higher-order contributions is introduced. The idea is similar to that in Ref. \cite{Melendez:2019izc}, but we do some simplifications for saving computation time.

In ChPT, physical quantities are calculated order by order. Each order provides a small factor $\epsilon=p/\Lambda_\chi$. For example, a physical quantity $X$ can be written as
\begin{align}
X=X_{\mathrm{ref}}\sum_{n=1}^{\infty}c_nQ^n,\label{equ:5}
\end{align}
where the dimensionless parameter $Q=\epsilon^2$, $c_n$ are dimensionless coefficients and $X_{\mathrm{ref}}$ is the natural size of $X$. We take $X_{\mathrm{ref}}$ equal to the LO value of $X$.

In practice, strict calculations in the higher orders are very complex because of a lot of unknown LECs and loop diagrams. Hence, the expansion of $X$ is truncated at a certain order and only the first few terms can be obtained. If $X$ is truncated at the order $k$, the theoretical prediction for $X$ is
\begin{align}
X^{\prime}=X_{\mathrm{ref}}\sum_{n=1}^{k}c_nQ^n, \label{equ:6}
\end{align}
where $k=1,\;2$, and 3 represent the LO, NLO and NNLO, respectively. The truncation error is 
\begin{align}
\Delta_k=X_{\mathrm{ref}}\sum_{n=k+1}^{\infty}c_nQ^n=X-X^{\prime}.\label{equ:7}
\end{align}
Before fitting the LECs, a nonzero and valid value of $\Delta_k$ needs be estimated first. For convenience, we estimate $X$ directly, but not $\Delta_k$.

According to hypothesis \ref{hy1}, the sequence $\{c_nQ^n\}$ is naively assumed to be a geometric sequence $\{a_0q^n\}$. Whether this assumption is reasonable or not depends on the final fitted results. This will be mentioned later. In this case, one can get the approximation,
\begin{align}
X=X_{\mathrm{ref}}\sum_{n=1}^{\infty}c_nQ^n\approx X_{\mathrm{ref}}\sum_{n=0}^{\infty}a_0q^n=X_{\mathrm{ref}}\frac{a_0}{1-q},\label{equ:8}
\end{align}
and the truncated error $\Delta_k$ is
\begin{align}
\Delta_k\approx X_{\mathrm{ref}}\frac{a_0}{1-q}-X^{\prime}.\label{equ:9}
\end{align}

In order to determine the parameters $a_0$ and $q$ in the geometric sequence, we define two cumulative sums sequences $\{S_k\}$ and $\{S^*_k\}$,
\begin{align}
S_k=\sum_{n=1}^{k}c_nQ^n,\quad S^*_k=\sum_{n=0}^{k}a_0q^n,\label{equ:10}
\end{align}
where the sequence $\{S_k\}$ can be regarded as a set of discrete data, and they can be calculated if a set of LECs related to $X$ is known. The cumulative sum $S_k^*$ of the geometric series is
\begin{align}
S^*(k)=\frac{a_0(1-q^{k+1})}{1-q}.\label{equ:11}
\end{align} 
The parameters $a_0$ and $q$ can be fitted by the least squares method.

For the NLO fit, only the LO and the NLO contributions of $F_S^\pi(t)$ can be calculated. Then only the NLO contributions of $\langle r^2\rangle_S^\pi$ and $c_S^\pi$ can be obtained according to Eqs. \eqref{equ:26} and \eqref{equ:27}. To fit $a_0$ and $q$ in Eq. \eqref{equ:11}, one needs to know at least two terms of $S^*(k)$. However, for $\langle r^2\rangle_S^\pi$ or $c_S^\pi$, the LO value is zero. We only know the NLO contribution at the NLO calculation. $a_0$ and $q$ can not be fitted at NLO. Hence, we only estimate $F_S^\pi(t)$ in Eqs. \eqref{equ:26} or \eqref{equ:27}. $K_{\ell4}$ form factors $f^{\prime}_s$ and $g^{\prime}_p$ have a similar property. We estimate them the same as $\langle r^2\rangle_S^\pi$ and $c_S^\pi$.

\subsection{Convergence}
Because the number of the observables is not large enough, the constraints on LECs are not very strong. In order to give more constraints on LECs, besides the observables $F_K/F_\pi$, $m_s/\hat m|_1$ and $m_s/\hat m|_2$, some other physical observables, i.e. $F_\pi$, $F_K$, $m_\pi$, $m_K$ and $m_\eta$ are also introduced separately. We find that not all of them have good convergence simultaneously. Sometimes, the two-flavor LECs $l_i^r\,(i=2,3)$ also have bad convergences. The NNLO values of these observables may be larger than the NLO ones. In other words, some of them may conflict with hypothesis \ref{hy1} in Sec. \ref{Sec:III}. Hence, we add the following new constraints to $\chi^2$ as Ref. \cite{Bijnens:2014lea},
\begin{align}
&f^\chi((m_\alpha^2)_6/m_\alpha^2/\Delta)\,(\alpha=\pi,K,\eta),\label{equ:320}\\ &f^\chi\Big(\Big(\frac{F_\alpha}{F_0}\Big)_6/\Delta\Big)\,(\alpha=\pi,K),\label{equ:321}\\ &f^\chi((l_i^r)_6/(l_i^r)_4/0.3)\,(i=2,3),\label{equ:32}
\end{align}
where the denominator $\Delta=0.1$, and $f^\chi(x)=2x^4/(1+x^2)$. If $f^\chi(x)=x^2$, it would be a normal $\chi^2$ distribution. When $x<1$, the chosen $f^\chi(x)$ is less than a normal $\chi^2$ distribution. When $x>1$, the chosen $f^\chi(x)$ is larger than a normal $\chi^2$ distribution. A smaller $x$ has a smaller influence on $\chi^2$ in Eq. \eqref{chi2}. Reference \cite{Bijnens:2014lea} only adopts the first equation.

\section{The results by Method I}\label{r1}
\subsection{NLO fitted $L_i^r$}\label{Sec:VI}

Table \ref{table1} presents the NLO fitted $L_i^r$. The results in the second column (fit 1) assume $L_4^r\equiv0$ and the other $L_i^r$ are left free in the fit. The results in the third column (fit 2) are obtained by a free fit. To compare with our results, the fourth (fifth) column lists the results in Ref. \cite{Bijnens:2014lea}, which are fitted at NLO (NNLO). The NLO fit averages three sets of results when $L_4=0,\pm0.3$. Both fit 1 and fit 2 are very close to the NNLO fit in Ref. \cite{Bijnens:2014lea}, but some of them are very different from its NLO fit. It indicates that the geometric series can give a good estimate for higher-order contributions. The estimation in Sec. \ref{Sec:IV} is valid. For fit 2, $L_4^r$ is small enough, and it is satisfied the large-$N_c$ limit. This is an assumption in Ref. \cite{Bijnens:2014lea}. Moreover, $2L_1^r-L_2^r$ and $L_6^r$ are also satisfied by the large-$N_c$ limit. They are also better than those in column 4. It means that the estimates from the higher order can not be ignored. They have a great influence on $L_i^r$ (especially $L_1^r$, $L_3^r$, $L_4^r$ and $L_6^r$), and the large-$N_c$ limit is satisfied automatically. Hence, we have a good reason to believe that contributions beyond the NNLO also have a great influence on $C_i^r$. When we calculate the NNLO LECs, the truncation errors need to be estimated too. Since fit 2 has no assumption about $L_4^r$ and its values are not very different from fit 1, we use it as the NLO results in this section.

\begin{table}[h]
	\caption{The NLO fitted $L_i^r$. The second and the third columns are our fitting results. The fourth and the fifth columns are the results in Ref. \cite{Bijnens:2014lea}, which is fitted at NLO and NNLO, respectively. The last line is $\chi^2$ and the degrees of freedom (d.o.f.). Fit 1: $L_4^r\equiv 0$. Fit 2: No assumption about $L_4^r$.}\label{table1}
	\begin{ruledtabular}
		\begin{tabular}{lcccc}
			LECs          &    Fit 1    &    Fit 2    & NLO fit \cite{Bijnens:2014lea} & NNLO fit \cite{Bijnens:2014lea} \\ \hline
			$10^3L_1^r$   & $0.42(05)$  & $0.44(05)$  &           1.0(1)           &            0.53(06)             \\
			$10^3L_2^r$   & $0.93(05)$  & $0.84(10)$  &           1.6(2)           &            0.81(04)             \\
			$10^3L_3^r$   & $-2.84(16)$ & $-2.84(16)$ &         $-$3.8(3)          &           $-$3.07(20)           \\
			$10^3L_4^r$   &  $\equiv$0  & $0.30(33)$  &           0.0(3)           &           $\equiv$0.3           \\
			$10^3L_5^r$   & $0.93(02)$  & $0.92(02)$  &           1.2(1)           &            1.01(06)             \\
			$10^3L_6^r$   & $0.18(05)$  & $0.22(08)$  &           0.0(4)           &            0.14(05)             \\
			$10^3L_7^r$   & $-0.22(12)$ & $-0.23(12)$ &         $-$0.3(2)          &           $-$0.34(09)           \\
			$10^3L_8^r$   & $0.44(10)$  & $0.44(10)$  &           0.5(5)           &            0.47(10)             \\
			$\chi^2$(d.o.f.) &   5.0(5)    &   4.2(4)    &            --(5)           &             1.0(9)
		\end{tabular}
	\end{ruledtabular}
\end{table}

The second to the fourth column in Table \ref{table2} shows the LO, the NLO and the higher-order contributions of the observables with fit 2 in Table \ref{table1}. The fifth column is the theoretical estimates. In order to see the convergence of these quantities obviously, the percentage of each order is defined,
\begin{align}
\mathrm{Pct}_{\mathrm{order}}=\Big|\frac{X_{\mathrm{order}}}{X_{\mathrm{th}}}\Big|\times 100\%,\label{equ:28}
\end{align}
where $X_{\mathrm{th}}$ is the theoretical estimate and the subscript ``order'' represents LO, NLO, and the higher order (HO). These percentages are shown in the parentheses from the second to the fourth columns in Table \ref{table2}. The experimental values are listed in the sixth column. It shows that all observables have good convergence. The $\chi_i^2=2.1$ from $\pi\,K$ scattering lengths ($a_0^{1/2}$ and $a_0^{3/2}$) give a dominant contribution to the total $\chi^2=4.2$. The main reason is that the LO contributions of these scattering lengths can not give good predictions. In other words, the LO contributions of $a_0^{1/2}$ and $a_0^{3/2}$ are very different from their experimental values. The contributions beyond the LO need to be large values, but the NLO contributions are not large enough. Hence, these two observables give about half of the total $\chi^2$. It seems that ChPT can not give a good prediction for these $\pi K$ scattering lengths. The convergences of $a_0^{1/2}$ and $a_0^{3/2}$ are bad and in conflict with hypothesis \ref{hy1} in Sec. \ref{Sec:III}. However, if they are not included in the fit, $L_4^r$ increases to $0.54\times10^{-3}$. This value conflicts with the large-$N_c$ limit. Hence, they are considered a necessity and will be included in the following calculations. The higher-order estimates of $f_s$, $g_p$, $a_0^0$, and $a_0^{1/2}$ are not very small. This is the reason why there are large deviations between the fourth column and the fifth column in Table \ref{table1}, such as $L_1^r$, $L_3^r$, $L_4^r$ and $L_6^r$.

\begin{table}[h]
	\caption{The convergences of the observables with fit 2 $L_i^r$ in Table \ref{table1}. The percentage $\mathrm{Pct}_{\mathrm{LO,NLO,HO}}$ is defined in Eq. \eqref{equ:28}. The last two columns are the theoretical estimates and experimental values, respectively.
  }\label{table2}
	\begin{ruledtabular}
		\begin{tabular}{lccccc}
			\multicolumn{1}{c}{Observables}&\multicolumn{1}{c}{LO$|\mathrm{Pct}_\mathrm{LO}$(\%)}&\multicolumn{1}{c}{NLO$|\mathrm{Pct}_\mathrm{NLO}$(\%)}&\multicolumn{1}{c}{HO$|\mathrm{Pct}_\mathrm{HO}$(\%)}&Theory&Experiment\\\hline
			$m_s/\hat m|_1$    &   25.8(94.0)    &   1.6(5.7)   &   0.1(0.4)    &   27.5   &   $27.3_{-1.3}^{+0.7}$   \\
			$m_s/\hat m|_2$    &   24.2(88.7)    &  2.7(10.0)   &   0.3(1.3)    &   27.3   &   $27.3_{-1.3}^{+0.7}$   \\
			$F_K/F_\pi$        &   1.000(83.4)   & 0.166(13.8)  &  0.033(2.8)   &  1.199   &  $1.199\pm 0.003$   \\
			$f_s$              &   3.782(66.2)   & 1.278(22.4)  &  0.652(11.4)  &  5.711   &  $5.712\pm 0.032$   \\
			$g_p$              &   3.782(76.7)   & 0.881(17.9)  &  0.268(5.4)   &  4.931   &  $4.958\pm 0.085$   \\
			$a_0^0$            &  0.1592(71.8)   & 0.0449(20.2) &  0.0176(7.9)  &  0.2217  &  $0.2196\pm 0.0034$  \\
			$10a_0^2$          & $-$0.455(104.8) &  0.022(5.0)  & $-$0.001(0.2) & $-$0.434 & $-0.444\pm 0.012$ \\
			$a_0^{1/2}m_\pi$   &   0.142(76.7)   & 0.033(17.8)  &  0.010(5.4)   &  0.185   &  $0.224\pm 0.022$   \\
			$10a_0^{3/2}m_\pi$ & $-$0.709(113.1) & 0.093(14.8)  & $-$0.011(1.7) & $-$0.627 & $-0.448\pm 0.077$ \\
			$F_S^\pi(0)/2B_0$  &   1.000(94.2)   &  0.058(5.4)  &  0.004(0.3)   &  1.061   &    --
		\end{tabular}
	\end{ruledtabular}
\end{table}

The convergences of $m_s/\hat m|_1$, $m_s/\hat m|_2$, and $F_K/F_\pi$ are already presented in Table \ref{table2}. If ChPT has good convergence, their numerators and denominators also need to be convergent separately. The NLO contributions of them are
\begin{align}
&\nonumber (m_\pi^2)_4=1.31\times 10^{-3}\mathrm{GeV}^2(7.2\%),\quad (m_K^2)_4=3.4\times 10^{-3}\mathrm{GeV}^2(1.4\%),\quad (m_\eta^2)_4=-1.16\times 10^{-2}\mathrm{GeV}^2(3.9\%),\\
&\Big(\frac{F_\pi}{F_0}\Big)_2+\Big(\frac{F_\pi}{F_0}\Big)_4=1+0.206,\quad\Big(\frac{F_K}{F_0}\Big)_2+\Big(\frac{F_K}{F_0}\Big)_4=1+0.372,\label{equ:29}
\end{align}
where the values in the parentheses in the first row are $(m^2_\alpha)_4/m^2_\alpha\;(\alpha=\pi,K,\eta)$. These observables also have satisfactory convergences.

\subsection{NNLO fitted $L_i^r$}\label{Sec:VII}

In the NNLO fit, the greatest difficulty is that 38 unknown LECs $C_i^r$ are involved. At present, we only find two methods can obtain all of their values. The latest results are in Refs. \cite{Bijnens:2014lea,Jiang:2015dba}, respectively. This subsection adopts these two sets of values first. We will estimate them in Sec. \ref{m2}.

There is a little difference from the NLO fit. Section \ref{Sec:VI} has mentioned that $\pi K$ scattering lengths $a_0^{1/2}$ and $a_0^{3/2}$ can not give good predictions and their NLO values are not large enough. Hence, the contributions beyond the NLO need large values. We assume that the truncation error should be small and the NNLO contribution should be large, because it is unnatural that the NNLO contribution is smaller than or approximately equal to the truncation error. It is difficult to estimate the values of $a_0^{1/2}$ and $a_0^{3/2}$ with the method in Sec. \ref{Sec:IV}. For example, the LO contribution of $a_0^{3/2}m_\pi$ is $-0.0709$ in Table \ref{table2}, but the experimental value is $-0.0448\pm0.0077$. If the NLO contribution has a small positive value, the NNLO contribution needs a larger positive value. Nevertheless, we have assumed the values at each order as a geometric sequence in Eq. \eqref{equ:8}. The first and the third term ($a_0$ and $a_0q^2$) in a geometric sequence have the same sign. The NNLO contribution should be negative too. To avoid this contradiction, we assume that $a_0^{1/2}$ and $a_0^{3/2}$ have good convergence except for the NNLO case. In this case, the truncation errors can be estimated according to only the LO and the NLO values, such as
\begin{align}
\Delta_{a_0^{1/2}}=(a_0^{1/2})_2\frac{q_1^3}{1-q_1},\quad q_1=\frac{(a_0^{1/2})_4}{(a_0^{1/2})_2}.\label{equ:30}
\end{align}
In this case, $\Delta_{a_0^{1/2}}$ is small if $q_1$ is small. The estimation of $\Delta_{a_0^{3/2}}$ is the similar to $\Delta_{a_0^{1/2}}$. For two-flavor LECs, Ref. \cite{Gasser:2007sg} gives the relation betweens $l_i^r$, $L_i^r$ and $C_i^r$ up to the NNLO. In the NNLO fit, $l_i^r(i=1,\ldots,4)$ are estimated first, which depend on the renormalized scale $\mu$. Then $\bar l_i$ are calculated by the estimates of $l_i^r$. 

\begin{table}[h]
	\caption{The NNLO fitted $L_i^r$. The results in the second and the fourth column use the $C_i^r$ in Refs. \cite{Bijnens:2014lea} and \cite{Jiang:2015dba}, respectively. The third and the fifth column are the relative deviations of $L_i^r$ [defined in Eq. \eqref{equ:31}]. The results in the last column are the NNLO fit in Ref. \cite{Bijnens:2014lea}.
  }\label{table3}
	\begin{ruledtabular}
		\begin{tabular}{lccccc}
			LECs          &      \multicolumn{2}{c}{Fit 3}       &      \multicolumn{2}{c}{Fit 4}       & NNLO fit\cite{Bijnens:2014lea} \\ \hline
			&   $L_i^r$   & $\mathrm{Pct}_{L_i^r}(\%)$ &   $L_i^r$   & $\mathrm{Pct}_{L_i^r}(\%)$ &                                \\\hline
			$10^3L_1^r$   & $0.37(05)$  &          20.5          & $0.44(05)$  &          0.4           &            0.53(06)            \\
			$10^3L_2^r$   & $0.74(04)$  &          13.4          & $0.35(04)$  &         140.5          &            0.81(04)            \\
			$10^3L_3^r$   & $-2.92(17)$ &          2.7           & $-2.16(16)$ &          31.8          &          $-$3.07(20)           \\
			$10^3L_4^r$   & $0.31(08)$  &          3.7           & $0.55(06)$  &          46.3          &          $\equiv$0.3           \\
			$10^3L_5^r$   & $1.01(03)$  &          8.1           & $1.03(02)$  &          10.4          &            1.01(06)            \\
			$10^3L_6^r$   & $0.29(04)$  &          21.8          & $0.14(05)$  &          55.6          &            0.14(05)            \\
			$10^3L_7^r$  & $-0.30(08)$ &          23.1          & $-0.05(06)$ &         322.8          &          $-$0.34(09)           \\
			$10^3L_8^r$   & $0.44(09)$  &          0.3           & $0.25(07)$  &          77.7          &            0.47(10)            \\
			$\chi^2$(d.o.f.) &     \multicolumn{2}{c}{14.7(9)}      &     \multicolumn{2}{c}{80.3(9)}      &            1.0(10)%
		\end{tabular}
	\end{ruledtabular}
\end{table}

The NNLO fitted $L_i^r$ are shown in Table \ref{table3}. Columns 2 and 4 use the $C_i^r$ in Refs. \cite{Bijnens:2014lea} and \cite{Jiang:2015dba}, respectively. Column 3 and 5 are the relative deviations of $L_i^r$
\begin{align}
\mathrm{Pct}_{L_i^r}=\Bigg|\frac{L^r_{i,\mathrm{NNLO}}-L^r_{i,\mathrm{NLO}}}{L^r_{i,\mathrm{NNLO}}}\Bigg|\times 100\%.\label{equ:31}
\end{align}
To compare with our results, the results in the last column are the NNLO fit in Ref. \cite{Bijnens:2014lea}.

For fit 3, $\chi^2/$d.o.f.$=14.7/9$ seems a little large. The main problem is that some $\mathrm{Pct}_{L_i^r}$ in column 3 are larger than 20\%, such as $\mathrm{Pct}_{L_1^r}$, $\mathrm{Pct}_{L_6^r}$ and $\mathrm{Pct}_{L_7^r}$. We consider these deviations are a little large. The value less than 20\% is acceptable. The results for the fit 4 are even worse. $\chi^2/$d.o.f.$=79.8/9$ is very large and most of $\mathrm{Pct}_{L_i^r}$ are larger than 20\%. Especially, the values of $L_2^r$ and $L_7^r$ are very different from their NLO fitted results. It indicates that these two sets of $C_i^r$ in the references can not fit the data well at NNLO. A new set of $C_i^r$ needs to be found. It needs to satisfy all the hypotheses in Sec. \ref{Sec:III}.

\section{Method II}\label{m2}

This section gives a new method to obtain a better set of $L_i^r$ and $C_i^r$ simultaneously. A part of processes in this method is similar to those in method I.

References \cite{Bijnens:2014lea,Bijnens:2011tb} estimate $C_i^r$ first, with a random-walk method in the parameter space of $C_i^r$, then they fit $L_i^r$ with the values of $C_i^r$. Although this method attempts to restrict the fitted values of $L_i^r$, some NNLO fitted values of $L_i^r$ deviate too much from their NLO fitted values (see Table \ref{table1}). For example, $L_1^r=0.53(06)\times10^{-3}$ and $L_2^r=0.81(04)\times10^{-3}$ at the NNLO fit, which are about half of their values at the NLO fit $L_1^r=1.0(1)\times10^{-3}$ and $L_2^r=1.6(2)\times10^{-3}$. We have attempted to determine $C_i^r$ by scattering points randomly in some possible parameter spaces, but we do not find a better set of $C_i^r$. In addition, we find that $C_i^r$ may not satisfy the large-$N_c$ limit. If we choose the nonzero $C_i^r$ in Ref. \cite{Jiang:2015dba}, but the zero $C_i^r$ are replaced by those in Ref. \cite{Bijnens:2014lea}, a much smaller $\chi^2$ can be found. Hence, we do not assume $C_i^r$ satisfies the large-$N_c$ limit. Any $C_i^r$ could have a not very small value.

Fit 2 in Table \ref{table1} gives a set of reasonable predictions in Table \ref{table2}. Its values are also close to the NNLO fit in Ref. \cite{Bijnens:2014lea}. Hence, we assume this set of $L_i^r$ is also close to the true values of $L_i^r$, and we take them as reference values. According to hypothesis \ref{hy2} in Sec. \ref{Sec:III}, The difference between the true value of $L_i^r$ and $L^r_{i,\mathrm{NLO}}$ is assumed to be less than 20\%, i.e.
\begin{align}
L_i^r\in[L^r_{i,\mathrm{NLO}}\times 80\%,~L^r_{i,\mathrm{NLO}}\times 120\%].\label{equ:33}
\end{align}
The value 20\% is a personal choice. We consider 20\% is large enough. If fit 2 in Table \ref{table1} is close enough to the true values of $L_i^r$, 20\% is even a little large for some $L_i^r$. We choose these wide ranges to cover the true values of $C_i^r$ as far as possible. These wide ranges will also lead to wide distributions of $C_i^r$. One can see how $C_i^r$ is clearly dependent on $L_i^r$.

The total number of $L_i^r$ and $C_i^r$ is $8+38=46$, which is much larger than the number of the observables 17. There are 29 redundant parameters. Theoretically, they can not be obtained exactly simultaneously. Hence, we expect that all $L_i^r$ are as precise as possible, but some $C_i^r$ could have large errors. We adopt the following steps to obtain all of them.
\begin{enumerate}
\item All $L_i^r$ are generated randomly according to a uniform distribution in the ranges in Eq. \eqref{equ:33}, because we do not know which values of $L_i^r$ are more possible. At this step, none of the $C_i^r$ are known, the calculation at NNLO does not start, and we also can not give the NNLO predictions for the observables. Theoretically, for any set of $L_i^r$, one could adjust the values of $C_i^r$ to give a good fit. In other words, before further calculations, we can not judge which $L_i^r$ in which range is more possible or not. A nonuniform distribution of $L_i^r$ may cause the final results of $L_i^r$ to be close to the peak of the distribution. However, at this step, we do not know where the peak is; hence, we choose uniform distribution. If the future study considers that the values of $L_i^r$ favor a particular distribution, one could modify uniform distribution to the new one.

It will show that if some calculations are done, a part of the generated $L_i^r$ could not lead to a reasonable set of $C_i^r$. Some sets of $L_i^r$ would conflict with the hypotheses in Sec. \ref{Sec:III}. They will be removed. The initial uniform distribution of a $L_i^r$ will change to a nonuniform one. We will use this new distribution to talk about $C_i^r$ in the following steps. In this step, $6.8\times10^{5}$ sets of $L_i^r$ are generated. This number is large enough to keep enough sets of $L_i^r$ at last.

\item \label{st} To avoid some unnecessary calculations, some sets of $L_i^r$ could be removed first by some simple constraints. We find that most sets of $L_i^r$ can not give a good prediction of $l_i^r$, i.e. the NLO theoretical values $(l_i^r)_4$ deviate too far away from their experimental values $(l_i^r)_{\mathrm{exp}}$. These sets of $L_i^r$ can not satisfy hypothesis \ref{hy1} in Sec. \ref{Sec:III} and they can be removed first. We use the following constraints,
\begin{align}
\Biggl|1-\frac{(l_i^r)_4}{(l_i^r)_{\mathrm{exp}}}\Biggl|\leq 0.8,\;(i=2,3).\label{equ:38}
\end{align}
The choice 0.8 is large enough and does not contradict with Eq. \eqref{equ:32}. This constraint seems very weak, but most sets of $L_i$ are removed because of this constraint. After this step, only about $6.8\times10^{4}$ sets of $L_i^r$ are left.

\item \label{sf} For a given set of $L_i^r$, the number of redundant parameters is $38-17=21$, which is still large. One can not obtain a unique set of $C_i^r$. The random-walk method \cite{Bijnens:2014lea,Bijnens:2011tb} may give a reasonable set of $C_i^r$, but the efficiency is low. It would take a very long time. Generally, different sets of $C_i^r$ may produce the same $\chi^2$. It is hard to confirm which one is more reasonable. Hence, we do not determine $C_i^r$ directly. On the other hand, there exist 17 observables. 17 functions of $C_i^r$ can be determined uniquely. Generally, not all of these functions are linear functions. Appendix \ref{app:A} provides a method to change them to linear ones called $\widetilde{C}_i$. For a given set of $L_i^r$, these $\widetilde{C}_i$ can be determined uniquely as method I. The remaining about $6.8\times10^{4}$ sets of $L_i^r$ lead to about $6.8\times10^{4}$ sets of $\widetilde{C}_i$.

The relation between $\widetilde{C}_i$ and $C^r_j$ in Appendix \ref{app:A} can be expressed as
\begin{align}
P_{ij}C^r_j=\widetilde{C}_i,\;(i=1\ldots 17),\label{equ:34}
\end{align}
where $P_{ij}$ is a coefficient matrix, $j$ is not continuous and its values can be found in Table \ref{table5}. To simplify this linear relation, one can multiply a suitable matrix $B$ on both sides of Eq. \eqref{equ:34}
\begin{align}
BPC^r=B\widetilde{C}\equiv\widetilde{C}',\label{equ:35}
\end{align}
where the matrix $BP$ is the reduced row echelon form of matrix $P$. Most $\widetilde{C}'_i$ are still linearly dependent on more than one $C^r_i$, but three $\widetilde{C}'_i$ are only related to $C_{14}^r$, $C_{15}^r$ and $C_{17}^r$, respectively. This is because three rows in the matrix $BP$ have only one nonzero element; the related three $C_i^r$ do not linearly combine with the others. In other words, $C_{14}^r$, $C_{15}^r$, and $C_{17}^r$ can be obtained directly, because $\widetilde{C}_i$ and $B$ are already known.

In this step, only the constraint in Eq. \eqref{equ:32} is used, because $A_{i}\,(i=1,\ldots,5)$ (defined in Appendix \ref{app:A}) can not be obtained separately. These $A_{i}$ are related to the NNLO values of $m_\alpha^2(\alpha=\pi,K,\eta)$ and $F_\alpha(\alpha=\pi,K)$. Hence, Eqs. \eqref{equ:320} and \eqref{equ:321} can not be calculated until now. These constraints are going to be adopted later.

\item \label{sa}There remain about $6.8\times10^{4}$ sets of $L_i^r$ and $\widetilde{C}_{i}$, but a lot of them give bad convergences of the observables. A typical three-flavor ChPT correction at the NLO, NNLO and NNNLO are $\sim 25\%$, $\sim 7\%$, and $\sim 1.5\%$, respectively \cite{Bijnens:2014lea}. For an observable $X$, except $a_0^{1/2}$, $a_0^{3/2}$ and $l^r_i (i=2,3)$, the following constraints are introduced
\begin{align}
\Biggl|\frac{(X)_{4}}{X}\Biggl|\times 100\%\leq30\%,\quad\Biggl|\frac{(X)_{6}}{X}\Biggl|\times 100\%\leq12\%, \quad\Biggl|\frac{(X)_{\mathrm{HO}}}{X}\Biggl|\times 100\%\leq7\%,\label{equ:39}
\end{align}
where the denominator $X$ is their theoretical estimates in Eq. \eqref{equ:8}. Because the typical corrections in each order are only a rough estimate, the upper bounds in Eq. \eqref{equ:39} are slightly more than the estimates.

$l^r_i (i=2,3)$ has been constrained by Eq. \eqref{equ:32}. For $\pi K$ scattering lengths $a_0^{1/2}$ and $a_0^{3/2}$, as discussed in Sec. \ref{Sec:VI}, they have a poor convergence. Hence, we assume they can have a larger NNLO contribution,
\begin{align}
\Biggl|\frac{(a_0^{1/2})_{6}}{a_0^{1/2}}\Biggl|\times 100\%\leq 25\%,\quad \Biggl|\frac{(a_0^{3/2})_{6}}{a_0^{3/2}}\Biggl|\times 100\%\leq 35\%,\label{equ:40}
\end{align}
where both denominators $a_0^{1/2}$ and $a_0^{3/2}$ are their theoretical estimates. Their constraints at NLO and the higher order are the same as those in Eq. \eqref{equ:39}. For $a_0^{3/2}$, its LO value is too small because of the discussion above Eq. \eqref{equ:30}.  Hence, the constraint at NNLO is looser than the other one.

After this step, some sets of $L_i^r$ and the relevant $\widetilde{C}_{i}$ do not satisfy Eqs. \eqref{equ:39} and \eqref{equ:40} and are removed.

\item Each remaining $\widetilde{C}_{i}$ has its own distribution. Some ranges could be very wide. However, in ChPT, the absolute values of $\widetilde{C}_{i}$ might not be very large. To constrain the ranges of $\widetilde{C}_{i}$, we take advantage of the values of $C_i^r$ in other references. Some references have estimated the values of $C_i^r$ (see Table \ref{Ci} in Appendix \ref{app:C}). With the help of these values, one can constrain the ranges of $\widetilde{C}_{i}$. However, not all of the results in the references are close to each other. Some results which are quite different from the others are excluded. The ranges of $C_{i}$ are chosen as
\begin{align}
C_{i}^r\in[{\bar C_{i}^r}-5\sigma_{C_{i}^r},\,{\bar C_{i}^r}+5\sigma_{C_{i}^r}],\label{ic}
\end{align}
where $\bar C_{i}^r$ are the mean value of $C_{i}^r$ in Table \ref{Ci} and $\sigma_{C_{i}^r}$ are their standard deviations. Several outliers are removed in the calculation. The numerical values can be found in Table \ref{cb}. The intervals chosen are $5\sigma$ wide. They are wide enough to cover nearly all values in the references. In addition, $3\sigma$-wide intervals can not give a large enough set of $C_{i}^r$ in our method (see the next step). Equation \eqref{ic} only gives the reasonable boundaries of $C_i^r$. If some more reasonable boundaries are found, it is not necessary to use all values in the literature.

The constraints for $m_\alpha^2(\alpha=\pi,K,\eta)$ and $F_\alpha(\alpha=\pi,K)$ in Eqs. \eqref{equ:320} and \eqref{equ:321} are replaced by the following constraints to constrain the ranges of $C^r_i$,
\begin{align}
\Bigg|\frac{(m_\alpha^2)_6}{m_\alpha^2}\Bigg|\leq 0.12\,(\alpha=\pi,K,\eta),\quad \Bigg|\Big(\frac{F_\alpha}{F_0}\Big)_6\bigg/\Big(\frac{F_\alpha}{F_0}\Big)\Bigg|\leq 0.12\,(\alpha=\pi,K),\label{mf}
\end{align}
where $F_\alpha/F_0$ is the theoretical estimate of the decay-constant ratios. Not all of $C^r_i$ satisfy these constraints. Relevant $\widetilde{C}_{i}$ and $L_i^r$ are also removed.

\item \label{sL} The distributions of most remaining $\widetilde{C}_i$ are similar to a normal distribution. Only a few of them have a little asymmetry. The mean values and standard deviations of $\widetilde{C}_i$ are regarded as their estimates and errors, respectively. To this step, 13114 sets of $L_i^r$ and $\widetilde{C}_i$ are left. This number is large enough in the statistical sense. To save computation time, we only use the mean values of $\widetilde{C}_i$ to estimate $C_{i}^r$, except for $C_{14}^r$, $C_{15}^r$, and $C_{17}^r$ (they have been obtained in Eq. \eqref{equ:35}).

Equations \eqref{equ:35} and \eqref{ic} describe a high-dimension range in the parameter space of $C_{i}^r$. Figure \ref{figure1} gives a low-dimensional example. The box in Fig. \ref{figure1} (a) gives the upper and the lower bounds of $C_i^r$ as Eq. \eqref{ic}. In the high-dimensional space, it is a hyperrectangle. The blue and the red planes are two possible constraints of $C_i^r$ for a given set of $\widetilde{C}_i$ as Eq. \eqref{equ:35}, if the matrix $B$ is $1\times3$ dimensional. For some sets of $\widetilde{C}_i$, the constraint plane may be similar to the blue one; for some other sets of $\widetilde{C}_i$, the constraint plane may be similar to the red one; and for the other sets of $\widetilde{C}_i$, the constraint plane can be the other possible cases. In the current situation, the box is $38-3=35$ dimensions and the plane is $35-17=18$ dimensions. The shape of the parameter space of $C_{i}^r$ (as the blue or the red range) is an 18-dimensional convex polyhedron. The geometric center of this parameter space is regarded as the estimates of $C_{i}^r$. In other words, we assume the possibilities are equal in the 18-dimensional convex polyhedron, because we have no reason which area is more possible.

We use the Monte-Carlo method to determine the centre, because we could not find an analytical method or a more effective numerical method. Considering the low-dimensional example in Fig. \ref{figure1} (a), the blue and the red convex polygons are laid flat in Fig. \ref{figure1} (b) and Fig. \ref{figure1} (c), respectively. One would scatter points in the blue/red range to determine its center, instead of the box in Fig. \ref{figure1} (a). However, we also could not find a method to scattering points only in this irregular range uniformly, especially in the high-dimensional space. One would scatter points in a larger rectangle range, such as $KLMN$ in Fig. \ref{figure1} (b) or $A_1  B_1 G_1 F_1$ in Fig. \ref{figure1} (c). The choice of the rectangle is not unique. This method works in the low-dimensional space, but in the high-dimensional space, the volume of the convex polyhedron might be very small compared with the volume of the smallest hyperrectangle. It would be very hard to scatter points in the convex polyhedron. Unfortunately, the values of $\widetilde{C}_i$ obtained above lead to this situation. Hence, instead of the hyperrectangle, we scatter points in a high-dimensional parallel polyhedron, such as $AGDI$ (or $BHEJ$) in Fig. \ref{figure1} (b) or $A_1 D_1 C_1 B_1$ (or $A_1 C_1 B_1 E_1$) in Fig. \ref{figure1} (c). The choice of the high-dimension parallel polyhedron is not unique too. There exist many possible cases. To increase the chances of searching points in the high-dimension convex polyhedron, the volume of the high-dimension parallel polyhedron is as small as possible. The volume of the smallest one might be less enough than the volume of the hyperrectangle. However, the total number of these high-dimension parallel polyhedrons is very large. Hence, we also use the Monte-Carlo method to choose a lot of different high-dimension parallel polyhedrons (not all of them), and then calculate all their volumes. Finally, we choose the smallest one. This method seems a little boring and takes a long time, but we find a small enough high-dimension parallel polyhedron at least. Now we can scatter points in this high-dimensional parallel polyhedron and pick up the points in the high-dimensional convex polyhedron. The mean values of the coordinates of these points are the estimates of $C_i^r$. The standard deviations of the coordinates of these points are regarded as the estimates of $\Delta C_{i}^r$. In the calculation, we find that the $3\sigma$-wide intervals are too narrow to generate points in the high-dimensional convex polyhedron. Hence, $5\sigma$-wide intervals are chosen in Eq. \eqref{ic}.

\begin{figure}
	\centering
	\includegraphics[width=1\linewidth]{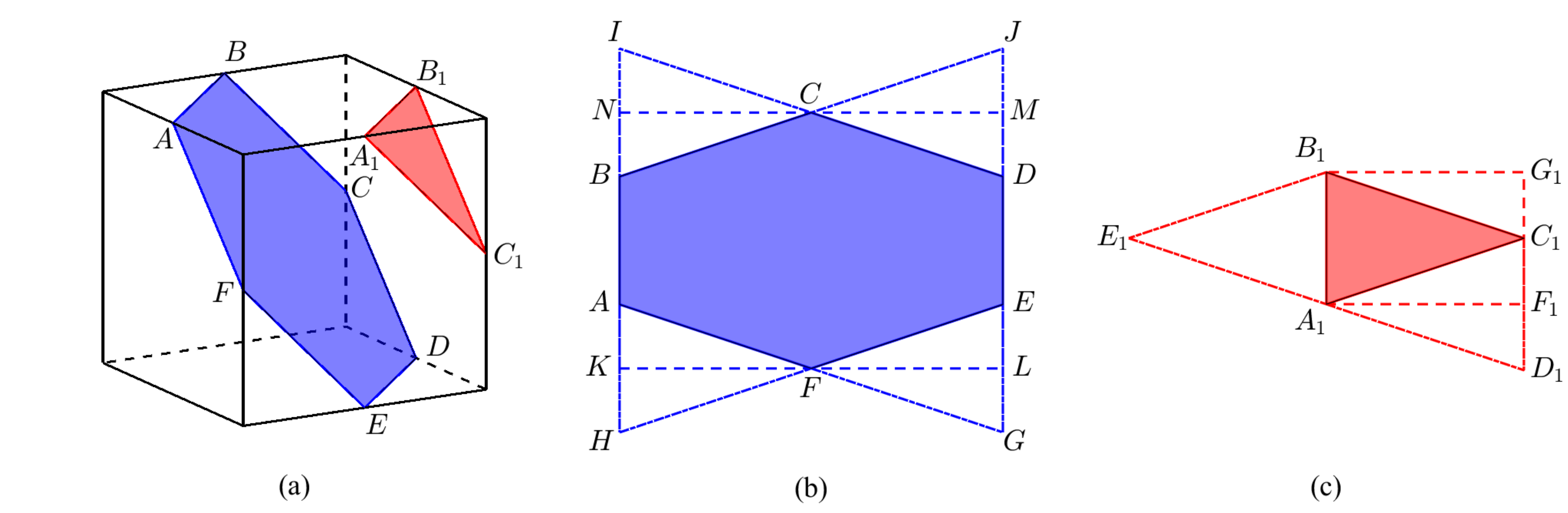}
	\caption{The possible parameter space of $C_i^r$ in three dimensions. The box in Fig. \ref{figure1} (a) gives the upper and the lower bounds of $C_i^r$ as Eq. \eqref{ic}. The blue and the red planes are two possible constraints of $C_i^r$ for a given set of $\widetilde{C}_i$ as Eq. \eqref{equ:35}, if the matrix $B$ is $1\times3$ dimensional. All possible values of $C_i^r$ are in the blue/red convex polygon. The blue and the red convex polygons are laid flat in Figs. \ref{figure1} (b) and \ref{figure1} (c), respectively.}
	\label{figure1}
\end{figure}

\item Using the estimates of $C_i^r$ obtained above, $L_i^r$ can be determined by method I.
\end{enumerate}

This method to obtain $L_i^r$ and $C_i^r$ with the truncation errors is general. It is only based on the hypotheses in Sec. \ref{Sec:III}. It does not depend on the number of observables, as long as the number of observables is larger than the number of $L_i^r$. In the present case, the number of observables (17) is small compared with the number of $C_i^r$ (38). Theoretically, if more observables are introduced, the values of $L_i^r$ are more precise. However, for $C_i^r$, besides giving more constraints of the existing $C_i^r$, the new observables may also introduce some new $C_i^r$. These new observables could also constrain the ranges of the new $C_i^r$. Sometimes, a new observable may give a new linearly independent $\widetilde{C}_i^r$. In this case, the rank of matrix $P$ in Eq. \eqref{equ:34} increases. A new constraint should be introduced and $C_i^r$ could be constrained in some narrower ranges. Otherwise, if the rank of matrix $P$ is not changed, this new observable has a little impact on $C_i^r$. It only leads to a more precise set of $\widetilde{C}_i$.

\section{The results by method II}\label{r2}
The estimates and errors of $\widetilde{C}_i$ are given in the Table \ref{table4}. If we randomly select a half of $\widetilde{C}_i$, the mean values and the standard deviations are unchanged. It indicates that the number of the samples is sufficient. Most of their relative errors are small enough, only $\Delta\widetilde{C}_i/\widetilde{C}_i\;(i=1,2,3,10)$ have large values. One could use this set of $\widetilde{C}_i$ to decide whether a part of $C_i^r$ are reasonable. A $\widetilde{C}_i$ is only related to a few $C_i^r$. For a particular research, sometimes it only needs a few $C_i^r$. If these $C_i^r$ are related to one or more $\widetilde{C}_i$, the values in Table \ref{table4} could decide to some extent whether these $C_i^r$ satisfy the observables discussed in this paper. If one knows more exact values of some $C_i^r$, some other $C_i^r$ can be obtained by these $\widetilde{C}_i^r$ too.

\begin{figure}
	\centering
	\includegraphics[width=1\linewidth]{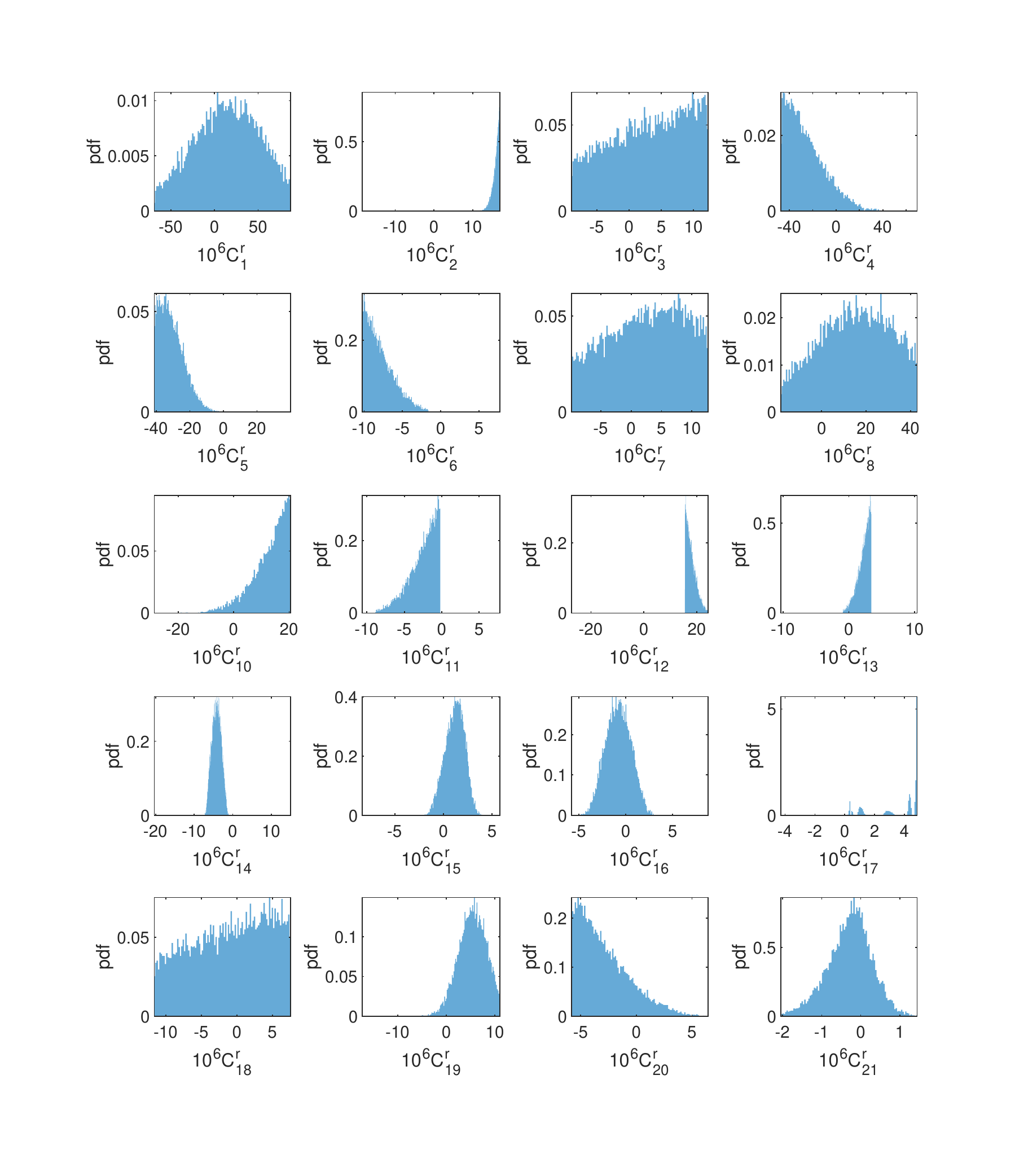}
  \caption{Distributions of the first part of $C_i^r$. The horizontal axis represents the value of $C_i^r$, the upper and the lower boundaries are given in Table \ref{cb}. The vertical axis represents the probability density function (pdf).}\label{figure2}
\end{figure}

\begin{figure}
	\centering
	\includegraphics[width=1\linewidth]{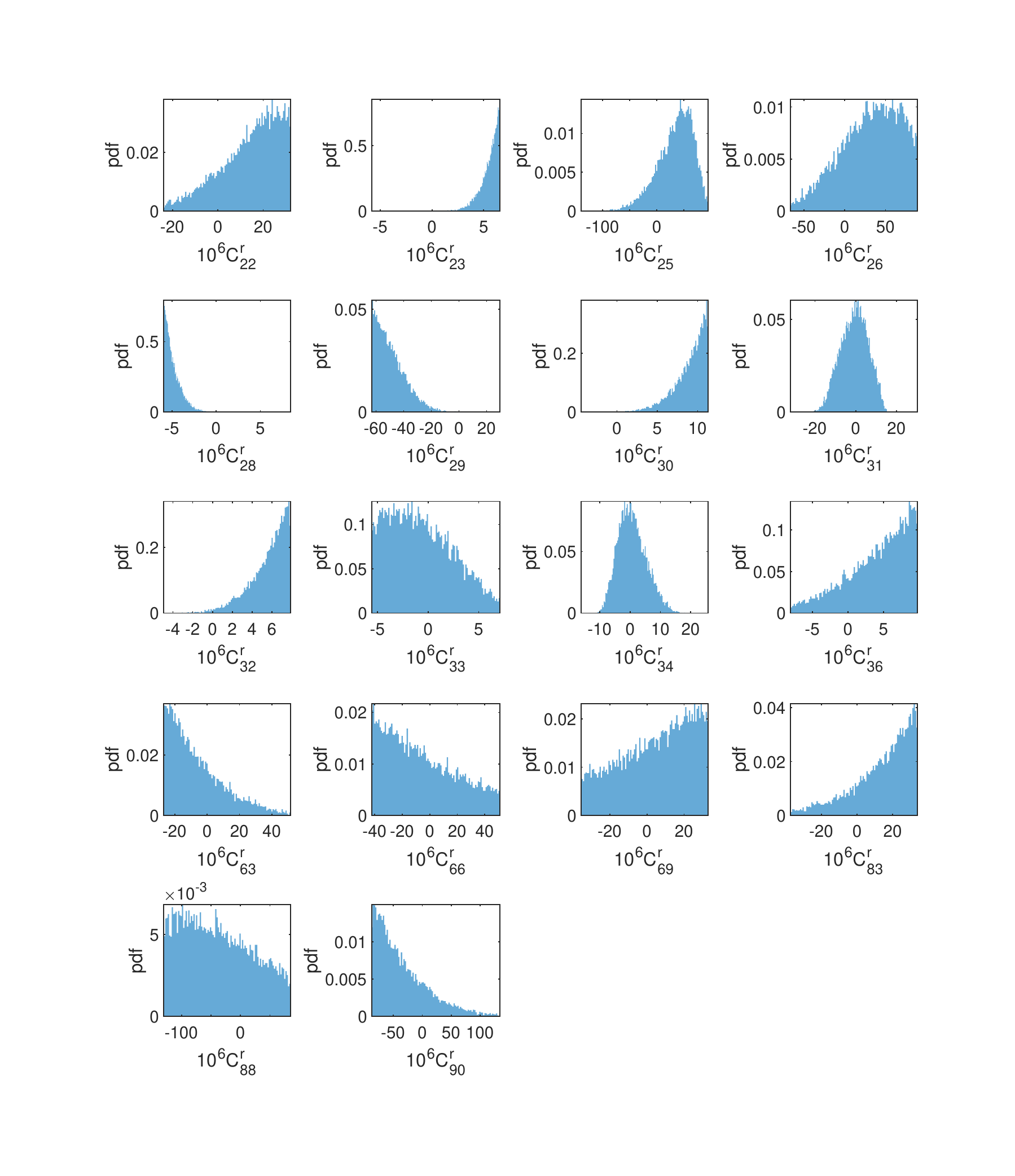}
	\caption{Distributions of the second part of $C_i^r$. The horizontal axis represents the value of $C_i^r$, the upper and the lower boundaries are given in Table \ref{cb}. The vertical axis represents the probability density function (pdf).}
	\label{figure3}
\end{figure}

\begin{table}[h]
	\caption{The estimates and errors of $\widetilde{C}_i$.}\label{table4}
	\begin{ruledtabular}
		\begin{tabular}{cccc}
			  $\widetilde{C}_i$   &   Values    &    $\widetilde{C}_i$     &   Values    \\ \hline
			  $\widetilde{C}_1$   & $0.02(12)$  &  $10\widetilde{C}_{10}$  & $-0.06(13)$ \\
			  $\widetilde{C}_2$   & $0.19(34)$  &   $\widetilde{C}_{11}$   & $0.24(02)$ \\
			$10^2\widetilde{C}_3$ & $-0.72(42)$ & $10^3\widetilde{C}_{12}$ & $-0.18(01)$ \\
			$10^2\widetilde{C}_4$ & $0.22(03)$  & $10^3\widetilde{C}_{13}$ & $1.02(44)$ \\
			 $10\widetilde{C}_5$  & $-0.16(02)$ & $10^4\widetilde{C}_{14}$ & $0.29(06)$ \\
			$10^3\widetilde{C}_6$ & $0.26(13)$  & $10^3\widetilde{C}_{15}$ & $-0.11(01)$ \\
			$10^2\widetilde{C}_7$ & $-0.42(12)$ & $10^4\widetilde{C}_{16}$ & $-0.56(06)$ \\
			 $10\widetilde{C}_8$  & $-0.45(09)$ & $10^4\widetilde{C}_{17}$ & $0.19(16)$ \\
			$10^2\widetilde{C}_9$ & $-0.99(11)$ &                          &
		\end{tabular}
	\end{ruledtabular}
\end{table}
The distributions of $C_i^r$ are shown in Fig. \ref{figure2} and Fig. \ref{figure3}. The upper and the lower boundaries in these figures are according to Eq. \eqref{ic}; their values are given in Table \ref{cb}. They show that most $C_i^r$ are dependent on the initial boundaries. $C_i^r\,(i=1,3,7,8,18,66,69,88)$ are dependent on both sides of the boundaries, and $C_i^r\,(i=2,4,5,6,10,17,20,22,23,26,28,29,30,32,33,36,63,83,90)$ are dependent on one side. To obtain a more precise value, one needs more reasonable constraints or more inputs. Eleven $C_{i}^r\,(i=11,\ldots,16,19,21,25,31,34)$ are nearly boundary independent and are more reasonable. $C_i^r\,(i=11,12,13)$ are very special. There exists a jump in the center. We solve the linear programming problem [Eqs. \eqref{equ:34} and \eqref{ic}] and find that does not there exist any solution in the other side of the jump. This is because the other boundaries of $C_i^r$ limit the solution. However, the jump is far away from both boundaries. These LECs are also considered boundary independent.

The values of $C_i^r$ are shown in Table \ref{table5}. Actually, in the calculation, the linear combinations $C^r_{63}-C^r_{83}+C^r_{88}/2$ and $C^r_{66}-C^r_{69}-C^r_{88}+C^r_{90}$ arise as a whole. We also present them in Table \ref{table5}. Our initial ranges of $C_i^r$ are wide and many sets of $C_i^r$ can give a small $\chi^2/$d.o.f. Hence, their errors seem large. There is a high probability that the true values of $C_i^r$ are in these ranges, especially the boundary-independent ones. Some results in Ref. \cite{Bijnens:2014lea} are marked with an asterisk, these values are very close to the original data on the website \cite{Bijnens2019} (less than $10^{-10}$). We guess these results meet the fitting boundaries. The symbol ``$\equiv0$'' in Column 4 and column 8 means these values are zeros in the large-$N_c$ limit. However, most intervals of $C_i^r\;(i=2,6,11,13,15,23)$ are far away from zero. It indicates that these LECs do not satisfy the large-$N_c$ limit. That is why the results in Ref. \cite{Jiang:2015dba} can not fit $L_i^r$ well by method I. 

\begin{table}[h]
	\caption{The values of $C_i^r$ in units of $10^{-6}$. The brackets ``['' and ``]'' mean the results are dependent on the lower and the upper boundaries, respectively. The parentheses ``('' and ``)'' mean the results are independent on the lower and the upper boundaries, respectively. The results with an asterisk mean the original data in the website \cite{Bijnens2019} are very close to those in Ref. \cite{Bijnens:2014lea} (less than $10^{-10}$). The symbol ``$\equiv0$'' for the results in Ref. \cite{Jiang:2015dba} means these values are zeros in the large-$N_c$ limits.}\label{table5}
	\begin{ruledtabular}
		\begin{tabular}{cccccccc}
			LECs     &     Results     & Ref. \cite{Bijnens:2014lea} & Ref. \cite{Jiang:2015dba} &                  LECs                  &    Results     & Ref. \cite{Bijnens:2014lea} & Ref. \cite{Jiang:2015dba} \\
			\hline
			 $C^r_1$   &   $14[37]$    &   12$^*$   & $25.33^{+0.60}_{-1.11}$  &              $C^r_{22}$               &   $14(13]$    &  9.0$^*$   & $-2.98^{+1.70}_{-2.21}$  \\
			 $C^r_2$   &    $16(1]$    &  3.0$^*$   &        $\equiv 0$        &              $C^r_{23}$               &  $5.6(0.9]$   & $-$1.0$^*$ &        $\equiv 0$        \\
			 $C^r_3$   &  $2.9[6.0]$   &  4.0$^*$   & $-0.43^{+0.09}_{-0.09}$  &              $C^r_{25}$               &   $34(33)$    & $-$11$^*$  & $-25.76^{-3.49}_{+5.02}$ \\
			 $C^r_4$   &   $-26[16)$   &   15$^*$   & $18.11^{+0.51}_{-0.85}$  &              $C^r_{26}$               &   $31(36]$    &     10     & $23.04^{+2.98}_{-4.59}$  \\
			 $C^r_5$   &   $-31[7)$    & $-$4.0$^*$ & $-10.88^{+0.85}_{-1.11}$ &              $C^r_{28}$               &  $-4.9[0.9)$  & $-$2.0$^*$ &  $1.53^{+0.00}_{-0.09}$  \\
			 $C^r_6$   &  $-7.9[1.8)$  & $-$4.0$^*$ &        $\equiv 0$        &              $C^r_{29}$               &   $-49[11)$   & $-$20$^*$  & $-8.42^{-1.79}_{+2.04}$  \\
			 $C^r_7$   &  $2.4[6.1]$   &  5.0$^*$   &        $\equiv 0$        &              $C^r_{30}$               &  $9.0(1.9]$   &  3.0$^*$   &  $3.15^{+0.09}_{-0.17}$  \\
			 $C^r_8$   &   $15[16]$    &   19$^*$   & $17.85^{-1.28}_{+1.36}$  &              $C^r_{31}$               & $-0.71(6.70)$ &  2.0$^*$   & $-3.91^{+0.60}_{-1.11}$  \\
			$C^r_{10}$ &    $13(6]$    &  $-$0.25   & $-5.53^{+0.43}_{-0.51}$  &              $C^r_{32}$               &  $5.6(1.9]$   &    1.7     &  $1.45^{-0.17}_{+0.26}$  \\
			$C^r_{11}$ &  $-2.6(1.8)$  & $-$4.0$^*$ &        $\equiv 0$        &              $C^r_{33}$               & $-0.69[3.12)$ &    0.82    & $-0.43^{-0.17}_{+0.43}$  \\
			$C^r_{12}$ &    $18(2)$    &   $-$2.8   & $-2.89^{+0.09}_{-0.09}$  &              $C^r_{34}$               & $0.68(4.67)$  &  7.0$^*$   &  $5.61^{-1.53}_{+2.47}$  \\
			$C^r_{13}$ &  $2.2(0.9)$   &    1.5     &        $\equiv 0$        &              $C^r_{36}$               &  $4.1(4.3]$   &  2.0$^*$   &        $\equiv 0$        \\
			$C^r_{14}$ &  $-4.2(1.2)$  & $-$1.0$^*$ & $-7.40^{+1.19}_{-1.79}$  &              $C^r_{63}$               & $-6.6[16.8)$  &     --     & $21.08^{-1.79}_{+2.13}$  \\
			$C^r_{15}$ &  $1.2(1.0)$   & $-$3.0$^*$ &        $\equiv 0$        &              $C^r_{66}$               & $-6.5[25.4]$  &     --     &  $6.80^{+0.34}_{-0.60}$  \\
			$C^r_{16}$ & $-0.81(1.34)$ &    3.2     &        $\equiv 0$        &              $C^r_{69}$               &  $4.6[19.0]$  &     --     &  $4.42^{+0.00}_{-0.09}$  \\
			$C^r_{17}$ &  $3.6(1.6]$   & $-$1.0$^*$ &  $1.45^{+0.09}_{-0.34}$  &              $C^r_{83}$               &   $14(16]$    &     --     & $-14.79^{+1.45}_{-1.87}$ \\
			$C^r_{18}$ &  $-1.1[5.4]$  &    0.63    & $-5.10^{+0.60}_{-0.77}$  &              $C^r_{88}$               &   $-38[59]$   &     --     & $-14.37^{-5.78}_{+7.91}$ \\
			$C^r_{19}$ &  $5.3(2.8)$   & $-$4.0$^*$ & $-2.30^{+0.77}_{-1.11}$  &              $C^r_{90}$               &   $-35[44)$   &     --     & $19.72^{-3.74}_{+4.68}$  \\
			$C^r_{20}$ &  $-2.9[2.3)$  &    1.0     &  $1.45^{-0.17}_{+0.26}$  &    $C^r_{63}-C^r_{83}+C^r_{88}/2$     &   $-39(33)$   &   $-$9.6   & $28.69^{-6.13}_{+7.96}$  \\
			$C^r_{21}$ & $-0.28(0.56)$ &  $-$0.48   & $-0.51^{+0.09}_{-0.09}$  & $C^r_{66}-C^r_{69}-C^r_{88}+C^r_{90}$ & $-7.9(78.9)$  &     50     & $36.47^{+2.38}_{-3.74}$
		\end{tabular}
	\end{ruledtabular}
\end{table}

The second column in Table \ref{table6} lists the results for $L_i^r$. Compared with Ref. \cite{Bijnens:2014lea}, $\chi^2/\mathrm{d.o.f.}=4.3/9$ is closer to 1. It means that the constraints in Sec. \ref{m2} relieve the overfitting problem. Convergences play an important role in the fit. For the normal distribution in Eq. \eqref{equ:33}, the initial standard deviation of ${L_i^r}=11.5\%$. Now, most relative deviations (in the third column) are obviously less than 11.5\%, except for $L_{2,3}^r$. Their values are a little large. We consider they are also acceptable. $\mathrm{Pct}_{L_2^r}$ and $\mathrm{Pct}_{L_3^r}$ in the third column in Table \ref{table6} are larger than the others. The main reason is that a set of $L_i^r$ containing small $L_2^r$ (large $L_3^r$) are much easier to be picked out in the step \ref{st} (\ref{sa}). For comparison, column 4 presents the average values of the remaining sets in step \ref{sL} in Sec. \ref{m2}, column 5 lists the relative deviations between column 4 and column 6. The values in column 2 and column 4 are close to each other. It seems that averaging $C_i^r$ or averaging $L_i^r$ first gives nearly no difference. These results are also not much different from fit 2 and the results in Ref. \cite{Bijnens:2014lea}. Because the results in column 2 are related to $C_i^r$ in Table \ref{table5}, we choose them as our $L_i^r$ results.

\begin{table}[h]
	\caption{The results for $L_i^r$. The second column is the final results of $L_i^r$. The fourth column is only a simple average of the value in step \ref{sL} in Sec. \ref{m2}. $\mathrm{Pct}_{L_i^r}$ in the second and the fourth column is defined in Eq. \eqref{equ:31}.}\label{table6}
	\begin{ruledtabular}
		\begin{tabular}{lcccccc}
			$L_i^r$       &   Results   & $\mathrm{Pct}_{L_i^r}(\%)$ &   Average   & $\mathrm{Pct}_{L_i^r}(\%)$ &    Fit 2    & Ref. \cite{Bijnens:2014lea} fit $p^6$ \\
			\hline
			$10^3L_1^r$   & $0.43(05)$  & 2.2  & $0.44(05)$  & 0.8  &  0.44(05)   &  0.53(06)   \\
			$10^3L_2^r$   & $0.74(04)$  & 14.0 & $0.77(05)$  & 10.2 &  0.84(10)   &  0.81(04)   \\
			$10^3L_3^r$   & $-2.47(17)$ & 15.0 & $-2.55(15)$ & 11.6 & $-$2.84(16) & $-$3.07(20) \\
			$10^3L_4^r$   & $0.33(08)$  & 9.3  & $0.30(03)$  & 2.2  &  0.30(33)   & $\equiv$0.3 \\
			$10^3L_5^r$   & $0.95(04)$  & 2.6  & $0.95(09)$  & 2.9  &  0.92(02)   &  1.01(06)   \\
			$10^3L_6^r$   & $0.20(03)$  & 9.9  & $0.21(02)$  & 8.8  &  0.22(08)   &  0.14(05)   \\
			$10^3L_7^r$   & $-0.23(08)$ & 2.2  & $-0.23(03)$ & 1.7  & $-$0.23(12) & $-$0.34(09) \\
			$10^3L_8^r$   & $0.42(09)$  & 5.4  & $0.42(04)$  & 4.7  &  0.44(10)   &  0.47(10)   \\
			$\chi^2$(d.o.f.) &   4.3(9)    &  --  &     --      &  --  &   4.2(4)    &   1.0(10)
		\end{tabular}
	\end{ruledtabular}
\end{table}               

In Table \ref{table7}, the values of the observables at each order and $\mathrm{Pct}_{\mathrm{order}}$ are listed. It can be seen that most observables have good convergence, except $a_0^{1/2}$ and $a_0^{3/2}$ at NNLO. The reason has been discussed in Sec. \ref{Sec:VI} and step \ref{sa} in Sec. \ref{m2}. Whether the higher-order values are really small or not requires a more reasonable analysis. It is beyond this work. The truncation errors in the fifth column are very small, all $\mathrm{Pct}_{\mathrm{HO}}$ are less than 4\%, except $l_2^r$. However, the absolute value of $l_2^r$ decreases order by order. It is not a contradiction. Now, with the hypothesises in Sec. \ref{Sec:III}, the first three problems in the Introduction are solved, and the cause of the last two problems is also found.

\begin{table}[h]
    \caption{The convergence of observables. The second to the fifth columns give the contributions and $\mathrm{Pct}_{\mathrm{order}}$ at each order. The theoretical values are given in the sixth column. The experimental values (inputs) are listed in the last column. $\bar{l}_i$ have been changed to $l_i^r$.}\label{table7}
	\begin{ruledtabular}
		\begin{tabular}{lcccccc}
			\multicolumn{1}{c}{physical quantities}&\multicolumn{1}{c}{LO$|\mathrm{Pct}_\mathrm{LO}$(\%)}&\multicolumn{1}{c}{NLO$|\mathrm{Pct}_\mathrm{NLO}$(\%)}&\multicolumn{1}{c}{NNLO$|\mathrm{Pct}_\mathrm{NNLO}$(\%)}&\multicolumn{1}{c}{HO$|\mathrm{Pct}_\mathrm{HO}$(\%)}&Theory&Inputs\\\hline
			$m_s/\hat m|_1$    &  $25.8(94.8)$   &   $2.0(7.2)$   &  $-1.1(4.0)$  &  $0.6(2.0)$   &  $27.3$  &   $27.3_{-1.3}^{+0.7}$   \\
			$m_s/\hat m|_2$    &  $24.2(88.7)$   &  $3.3(12.2)$   &  $-0.8(2.8)$  &  $0.5(1.9)$   &  $27.3$  &   $27.3_{-1.3}^{+0.7}$   \\
			$F_K/F_\pi$        &  $1.000(83.4)$  & $0.169(14.1)$  & $0.023(1.9)$  & $0.007(0.6)$  & $1.199$  &  $1.199\pm 0.003$   \\
			$f_s$              &  $3.782(66.2)$  & $1.322(23.1)$  & $0.371(6.5)$  & $0.235(4.1)$  & $5.709$  &   $5.712\pm 0.032$      \\
			$g_p$              &  $3.782(76.7)$  & $0.776(15.7)$  & $0.366(7.4)$  & $0.007(0.1)$  & $4.931$  &  $4.958\pm 0.085$     \\
			$a_0^0$            & $0.1592(72.5)$  & $0.0453(20.6)$ & $0.0098(4.5)$ & $0.0053(2.4)$ & $0.2196$ &   $0.2196\pm 0.0034$   \\
			$10a_0^2$          & $-0.455(103.8)$ &  $0.022(5.0)$  & $-0.010(2.2)$ & $0.005(1.1)$  & $-0.438$ & $-0.444\pm 0.012$ \\
			$a_0^{1/2}m_\pi$   &  $0.142(62.6)$  & $0.033(14.6)$  & $0.049(21.7)$ & $0.002(1.0)$  & $0.226$  &  $0.224\pm 0.022$    \\
			$10a_0^{3/2}m_\pi$ & $-0.709(150.2)$ & $0.094(19.8)$  & $0.142(30.1)$ & $0.001(0.3)$  & $-0.472$ & $-0.448\pm 0.077$  \\
			$F_S^\pi(0)/2B_0$  &  $1.000(98.1)$  &  $0.019(1.9)$  & $0.000(0.0)$  & $0.000(0.0)$  & $1.019$  &    --    \\
			$10^3l_1^r$        &       --        &  $-3.2(81.5)$  & $-0.6(15.1)$  &  $-0.1(3.4)$  &  $-4.0$  &   $-4.0\pm0.6$  \\
			$10^3l_2^r$        &       --        &  $3.0(145.6)$  & $-1.3(66.5)$  &  $0.4(20.8)$  &  $2.0$   &   $1.9\pm0.2$  \\
			$10^3l_3^r$        &       --        &  $0.2(102.5)$  &  $-0.0(2.6)$  &  $0.0(0.1)$   &  $0.2$   &   $0.3\pm1.1$  \\
			$10^3l_4^r$        &       --        &  $6.3(96.8)$   &  $0.2(3.1)$   &  $0.0(0.1)$   &  $6.6$   &   $6.2\pm1.3$	\end{tabular}
	\end{ruledtabular}
\end{table}

\section{summary and discussion}\label{sum}

In this paper, we have computed the NLO and the NNLO LECs for pseudoscalar mesons with a new method (method II). The results are present in Table \ref{table6} and Table \ref{table5}, respectively. The truncation errors are considered in the computation with some hypotheses in Sec. \ref{Sec:III}; i.e., the theoretical values of observables are satisfied with the convergence in ChPT, all $L_i^r$ are stable and $C_i^r$ are consistent with those in the other references. The results nearly satisfy all these hypotheses, and all random processes are repeated several times. The results are nearly unchanged. They are reasonable in statistics. Some linear combinations of $C_i^r$ called $\widetilde{C}_i^r$ are also given. Their relative errors are less than $C_i^r$ and their values are more reliable. If one knows more exact values of some $C_i^r$, some other $C_i^r$ can be obtained by these $\widetilde{C}_i^r$.

First, a modified global fit method is used to obtain $L_i^r$. If they are only fitted at NLO, the results are very close to the NNLO fitted results in Ref. \cite{Bijnens:2014lea}. It indicates that the higher-order estimates have a good prediction. The estimation is reasonable. However, some $L_i^r$ deviate from the NLO fitting ones in Ref. \cite{Bijnens:2014lea} too much. The main reason is that the higher-order estimates of $f_s$, $g_p$, $a_0^0$ and $a_0^{1/2}$ are not very small; i.e., the truncation errors can make a great impact on the values of the lower order LECs. The $\pi K$ scattering lengths $a_0^{1/2}$ and $a_0^{3/2}$ can not be fitted well, because their LO contributions can not give a good prediction, and the NLO contributions tend to be small. For the NLO fit, all the theoretical values of observables have good convergence. However, at NNLO, we have tried two sets of $C_i^r$ in the references, but the results are not very good. 

Later, we use a new method to obtain both $L_i^r$ and $C_i^r$. The idea is that the linearly independent combinations of $C_i^r$ ($\widetilde{C}_i$) are obtained first, and then $C_i^r$ are estimated with the Monte-Carlo method. Finally, $L_i^r$ are fitted by these $C_i^r$. Some $C_i^r$ are dependent on the initial boundaries. In order to obtain more precise values, these $C_i^r$ need more information to narrow the boundaries. The other $C_i^r$ are boundary independent and can be limited to some reliable intervals. The relative errors between the NLO fitting results and the one by this method are all small. All observables have good convergence, except $a_0^{1/2}$ and $a_0^{3/2}$. For $a_0^{1/2}$ and $a_0^{3/2}$, we assume that their contributions beyond NNLO are small and their NNLO contributions are large, because their NLO values are not large enough. Whether this assumption is correct or not needs a more reasonable estimate beyond the NNLO according to ChPT.

Some constraints in this paper are very weak, such as Eqs. \eqref{equ:33}, \eqref{equ:38}, \eqref{equ:39}, \eqref{ic}, and \eqref{mf}, because we want to cover the true value as far as possible. Hence, some error ranges of the NNLO LECs are large. If another method can introduce more restrictive constraints, their error ranges may be narrower. However, the values of $L_i^r$ are more reliable. Their NLO and NNLO fitting results are very closed. One could use these $L_i^r$ for further calculations directly. However, the ranges of $C_i^r$ are larger. They could only give rough estimates at NNLO. The estimates could be treated as references to judge whether the NNLO contributions are small or large. We hope that this new method not only determines the LECs in ChPT for mesons, but it will also generalize to ChPT for baryons and another effective field theory in the future. Other effective field theories should also contain a lot of LECs, especially at the high order. If one fits the low-order LECs, the fit at the low order should also give a reliable set of LECs with truncation errors, even without the higher calculations. It would save a lot of effort. Some new predictions could be given with these fitting LECs. Some choices in this paper are also personal favorites. For example, the three hypotheses in Sec. \ref{Sec:III} are very rough, for some special cases, stricter or looser conditions might be introduced. We also choose uniform distribution in Eq. \eqref{equ:33}, but the strict distribution would be very complex. These need to be further studied in the future too.

\section{acknowledgements}

We thank Professor Johan Bijnens for some help with $K_{\ell4}$ program. We also thank Dr. Da-Bin Lin for providing a powerful computer. Yang and Jiang thank Dr. Sarah Wesolowski for giving a lot of helpful comments. This work was supported by the National Science Foundation of China (NSFC) under Grant No. 11565004, Guangxi Science Foundation under Grants No. 2018GXNSFAA281180 and No. 2017AD22006, and the high-performance computing platform of Guangxi University.
\appendix

\section{The linear combinations of $C_i^r$}\label{app:A}

In Sec. \ref{m2}, most $C_i^r$ have not been calculated separately, because the number of redundant parameters is very large. In this appendix, some $\widetilde{C}_i$ will be defined. They are linear combinations of $C_i^r$. These $\widetilde{C}_i$ can be calculated separately.

Generally, the NNLO contribution of some observables $X_j$ can be separated into two parts, one part is proportional to $C_i^r$ ($X_{j,C}$) and the other part is related to $L_i^r$ ($X_{j,L}$)
\begin{align}
X_j=X_{j,L}+X_{j,C}=X_{j,L}+d_jA_j,\label{equ:A1}
\end{align}
where different $j$ denotes different observables which will be discussed below, $d_j$ are possible dimensional parameters, and $A_j$ are dimensionless coefficients. $d_j$ are independent of $C_i^r$, but $A_j$ are dependent on $C_i^r$. In this section, the discussion is only about $X_{j,C}$,  so we will not go into detail below.

For meson masses and decay constants, $j=1\ldots5$ denote $m_\pi^2,m_K^2,m_\eta^2,F_\pi/F_0,F_K/F_0$, respectively; $d_{1,2,3}=m_\pi^6/F_\pi^4$, $d_{4,5}=m_\pi^4/F_\pi^4$ and $A_j$ are
\begin{align}
\nonumber
A_1=&48\,C^r_{19}-16\,C^r_{14}-16\,C^r_{17}-32\,C^r_{12}+32\,C^r_{31}-C^r_{15}\,\left(32\,a^2+16\right)-C^r_{13}\,\left(64\,a^2+32\right)\nonumber\\
&+C^r_{32}\,\left(64\,a^2+32\right)+C^r_{20}\,\left(64\,a^4+80\right)-C^r_{16}\,\left(64\,a^4-64\,a^2+48\right)+48\,C^r_{21}\,{\left(2\,a^2+1\right)}^2,\label{equ:A5}\\
A_2=&32\,C^r_{31}\,a^6-32\,C^r_{12}\,a^6-16\,C^r_{14}\,a^2\,\left(2\,a^4-2\,a^2+1\right)+48\,C^r_{19}\,a^2\,\left(2\,a^4-2\,a^2+1\right)\nonumber\\
&-16\,C^r_{16}\,a^2\,\left(4\,a^4-4\,a^2+3\right)+16\,C^r_{20}\,a^2\,\left(8\,a^4-2\,a^2+3\right)+48\,C^r_{21}\,a^2\,{\left(2\,a^2+1\right)}^2\nonumber\\
&-32\,C^r_{13}\,a^4\,\left(2\,a^2+1\right)-16\,C^r_{15}\,a^4\,\left(2\,a^2+1\right)-16\,C^r_{17}\,a^2\,\left(2\,a^2-1\right)+32\,C^r_{32}\,a^4\,\left(2\,a^2+1\right),\label{equ:A6}\\
A_3=&C^r_{20}\,\left(256\,a^6-192\,a^4+64\,a^2+16\right)+C^r_{19}\,\left(256\,a^6-384\,a^4+192\,a^2-16\right)\nonumber\\
&-C^r_{16}\,\left(\frac{256\,a^6}{3}-\frac{320\,a^4}{3}+\frac{256\,a^2}{3}-16\right)-C^r_{32}\,\left(-\frac{512\,a^6}{3}+\frac{256\,a^4}{3}+\frac{64\,a^2}{3}-32\right)\nonumber\\
&+C^r_{31}\,\left(\frac{512\,a^6}{3}-256\,a^4+128\,a^2-\frac{32}{3}\right)-C^r_{14}\,\left(\frac{512\,a^6}{9}-\frac{640\,a^4}{9}+\frac{320\,a^2}{9}-\frac{16}{3}\right)\nonumber\\
&-C^r_{17}\,\left(\frac{512\,a^6}{9}-\frac{640\,a^4}{9}+\frac{320\,a^2}{9}-\frac{16}{3}\right)-\frac{32\,C^r_{12}\,{\left(4\,a^2-1\right)}^3}{27}\nonumber\\
&-\frac{32\,C^r_{13}\,\left(2\,a^2+1\right)\,{\left(4\,a^2-1\right)}^2}{9}-\frac{16\,C^r_{15}\,\left(2\,a^2+1\right)\,{\left(4\,a^2-1\right)}^2}{9}\nonumber\\
&+16\,C^r_{21}\,{\left(2\,a^2+1\right)}^2\,\left(4\,a^2-1\right)-\frac{128\,C^r_{18}\,{\left(a^2-1\right)}^2\,\left(4\,a^2-1\right)}{9}+\frac{512\,C^r_{33}\,a^2\,{\left(a^2-1\right)}^2}{3},\label{equ:A7}\\
A_4=&8\,C^r_{14}+8\,C^r_{17}+C^r_{15}\,\left(16\,a^2+8\right)+C^r_{16}\,\left(32\,a^4-32\,a^2+24\right),\label{equ:A8}\\
A_5=&C^r_{17}\,\left(16\,a^2-8\right)+C^r_{14}\,\left(16\,a^4-16\,a^2+8\right)+C^r_{16}\,\left(32\,a^4-32\,a^2+24\right)+8\,C^r_{15}\,a^2\,\left(2\,a^2+1\right),\label{equ:A9}
\end{align}
where $a=m_K/m_\pi$. For $m_s/\hat m|_1$, $m_s/\hat m|_2$ and $F_K/F_\pi$, the NNLO order contributions related to $C_i^r$ are $(m_\pi^4/F_\pi^4)\widetilde{C}_i$ $(i=1,2,3)$ respectively, where
\begin{align}
\widetilde{C}_1=&\frac{m_K^2}{m_\pi^2}A_1-A_2,\label{equ:A2}\\
\widetilde{C}_2=&\frac{m_\eta^2}{m_\pi^2}A_1-A_3,\label{equ:A3}\\
\widetilde{C}_3=&A_5-A_4.\label{equ:A4}
\end{align}

For $K_{\ell4}$ form factors, the NNLO contributions of $f_s$ and $f_s^{\prime}$ can be written as \cite{Amoros:2000mc}
\begin{align}
&\hspace{-20pt}F(s_\pi,s_{\ell}=0,\cos\theta=0)_6=F_{6,L}+F_{6,C}=F_{6,L}+\frac{1}{F_\pi^4}(A_6s_\pi^2+A_7s_\pi m_\pi^2+A_8 m_\pi^4).\label{equ:A10}\\
A_6=&4\,C^r_{3}-64\,C^r_{2}-14\,C^r_{1}+20\,C^r_{4},\label{equ:A16}\\\nonumber
A_7=&C^r_{10}\,\left(4\,a^2+4\right)+C^r_{5}\,\left(4\,a^2+16\right)+C^r_{8}\,\left(16\,a^2+4\right)+C^r_{12}\,\left(12\,a^2-8\right)\\\nonumber
&+C^r_{22}\,\left(4\,a^2+8\right)+C^r_{11}\,\left(16\,a^2+8\right)+C^r_{4}\,\left(4\,a^2-32\right)-C^r_{23}\,\left(8\,a^2+16\right)\\\nonumber
&+C^r_{25}\,\left(8\,a^2+16\right)+C^r_{1}\,\left(10\,a^2+48\right)+C^r_{6}\,\left(40\,a^2+20\right)+C^r_{7}\,\left(32\,a^2+32\right)\\
&+C^r_{13}\,\left(48\,a^2-40\right)+C^r_{2}\,\left(32\,a^2+192\right)+4\,C^r_{3}\,a^2,\label{equ:A17}\\\nonumber
A_8=&128\,C^r_{16}+128\,C^r_{28}+C^r_{5}\,\left(4\,a^4-32\right)-C^r_{1}\,\left(16\,a^2+32\right)-C^r_{14}\,\left(16\,a^2-32\right)\\\nonumber
&-C^r_{26}\,\left(16\,a^2-32\right)+C^r_{15}\,\left(80\,a^2+8\right)+C^r_{17}\,\left(64\,a^2-48\right)-C^r_{7}\,\left(64\,a^2+64\right)\\\nonumber
&-C^r_{2}\,\left(64\,a^2+128\right)-C^r_{6}\,\left(-8\,a^4+60\,a^2+32\right)-C^r_{12}\,\left(12\,a^4-24\,a^2+64\right)\\\nonumber
&-C^r_{13}\,\left(16\,a^4+72\,a^2+64\right)-28\,C^r_{8}\,a^2-16\,C^r_{25}\,a^2-32\,C^r_{29}\,a^2-64\,C^r_{30}\,a^2\\\nonumber
&-8\,C^r_{34}\,a^4-32\,C^r_{36}\,a^2+4\,C^r_{10}\,a^2\,\left(a^2+1\right)+8\,C^r_{23}\,a^2\,\left(a^2-2\right)+4\,C^r_{22}\,a^2\,\left(a^2-6\right)\\
&+8\,C^r_{11}\,a^2\,\left(2\,a^2+1\right).\label{equ:A18}
\end{align}
$f'_s$ can be calculated numerically
\begin{align}
f'_s=4m_\pi^2\frac{F(s'_\pi)-F(s_\pi)}{s'_\pi-s_\pi},\label{equ:A11}
\end{align} 
where $s_\pi=(2m_\pi+0.001 \mathrm{MeV})^2$ and $s'_\pi=(293\mathrm{MeV})^2$ are around the threshold. The two observables $f_s$ and $f'_s$ are related to two independent linear combinations
\begin{align}
\widetilde{C}_4&=A_6-\frac{m_\pi^4}{s_\pi s'_\pi}A_8,\label{equ:A12}\\
\widetilde{C}_5&=A_7+\frac{m_\pi^2(s_\pi+s'_\pi)}{s_\pi s'_\pi}A_8.\label{equ:A13}
\end{align}
The discussion for $g_p$ and $g^{\prime}_p$ is similar to $f_s$ and $f_s^{\prime}$. The parameters $A_{9,10,11}$ and the independent linear combinations $\widetilde{C}_{6,7}$ are
\begin{align}
A_9=&4\,C^r_{3}-2\,C^r_{1}+2\,C^r_{4}+3\,C^r_{66}-3\,C^r_{69}-3\,C^r_{88}+3\,C^r_{90},\label{equ:A19}\\\nonumber
A_{10}=&C^r_{10}\,\left(4\,a^2+4\right)-C^r_{6}\,\left(8\,a^2+4\right)-4\,C^r_{8}-C^r_{4}\,\left(8\,a^2+8\right)-C^r_{12}\,\left(4\,a^2+16\right)\\\nonumber
&+C^r_{22}\,\left(8\,a^2+4\right)+C^r_{11}\,\left(16\,a^2+8\right)-C^r_{25}\,\left(8\,a^2+4\right)+C^r_{63}\,\left(4\,a^2-4\right)\\\nonumber
&-C^r_{66}\,\left(2\,a^2+4\right)+C^r_{69}\,\left(2\,a^2+4\right)-C^r_{13}\,\left(48\,a^2+24\right)-C^r_{83}\,\left(4\,a^2-4\right)\\
&+C^r_{88}\,\left(4\,a^2+2\right)-C^r_{90}\,\left(2\,a^2+4\right)-2\,C^r_{1}\,a^2+4\,C^r_{3}\,a^2-4\,C^r_{5}\,a^2,\label{equ:A20}\\\nonumber
A_{11}=&16\,C^r_{17}+C^r_{15}\,\left(16\,a^2+8\right)+C^r_{66}\,\left(4\,a^2-a^4\right)+C^r_{90}\,\left(4\,a^2-a^4\right)-4\,C^r_{5}\,a^4-4\,C^r_{8}\,a^2\\\nonumber
&-28\,C^r_{12}\,a^4+16\,C^r_{14}\,a^2-20\,C^r_{22}\,a^2+20\,C^r_{25}\,a^2+16\,C^r_{26}\,a^2-32\,C^r_{29}\,a^2-8\,C^r_{34}\,a^4\\\nonumber
&-C^r_{88}\,\left(a^4+2\,a^2\right)-2\,C^r_{4}\,a^2\,\left(a^2-4\right)+4\,C^r_{10}\,a^2\,\left(a^2+1\right)-4\,C^r_{63}\,a^2\,\left(a^2-1\right)\\\nonumber
&+C^r_{69}\,a^2\,\left(a^2-4\right)+4\,C^r_{83}\,a^2\,\left(a^2-1\right)-4\,C^r_{6}\,a^2\,\left(2\,a^2+1\right)+8\,C^r_{11}\,a^2\,\left(2\,a^2+1\right)\\
&-24\,C^r_{13}\,a^2\,\left(2\,a^2+1\right).\label{equ:A21}\\
\widetilde{C}_6=&A_9-\frac{m_\pi^4}{s_\pi s'_\pi}A_{11},\label{equ:A14}\\
\widetilde{C}_7=&A_{10}+\frac{m_\pi^2(s_\pi+s'_\pi)}{s_\pi s'_\pi}A_{11}.\label{equ:A15}
\end{align}

The NNLO contribution of the $\pi\pi$ scattering amplitude is related to $A(s,t,u)$ and $A(t,u,s)=A(u,s,t)$, where $s=4m_\pi^2$, $t=0$ and $u=0$. They can be written as Eq. \eqref{equ:A1} \cite{Bijnens:2004eu},
\begin{align}
A_{12}=&192\,C^r_{3}-128\,C^r_{2}-64\,C^r_{1}+384\,C^r_{4}+32\,C^r_{5}+64\,C^r_{7}+32\,C^r_{8}+32\,C^r_{10}-96\,C^r_{12}\nonumber\\
&+64\,C^r_{14}+128\,C^r_{16}+64\,C^r_{17}+96\,C^r_{19}-128\,C^r_{22}-128\,C^r_{23}+192\,C^r_{25}+64\,C^r_{26}\nonumber\\
&+128\,C^r_{28}-192\,C^r_{29}-128\,C^r_{30}+96\,C^r_{31}+C^r_{6}\,\left(64\,a^2+32\right)+C^r_{11}\,\left(64\,a^2+32\right)\nonumber\\
&+C^r_{15}\,\left(64\,a^2+96\right)-C^r_{13}\,\left(64\,a^2+160\right)+C^r_{20}\,\left(64\,a^2+160\right)+C^r_{32}\,\left(64\,a^2+160\right)\nonumber\\
&+C^r_{21}\,\left(384\,a^2+192\right),\label{equ:A26}\\
A_{13}=&64\,C^r_{1}+128\,C^r_{2}-64\,C^r_{3}-128\,C^r_{4}+32\,C^r_{5}+64\,C^r_{7}+32\,C^r_{8}+32\,C^r_{10}-96\,C^r_{12}\nonumber\\
&-64\,C^r_{14}-128\,C^r_{16}-64\,C^r_{17}+96\,C^r_{19}+128\,C^r_{22}+128\,C^r_{23}-64\,C^r_{25}-64\,C^r_{26}\nonumber\\
&-128\,C^r_{28}+64\,C^r_{29}+96\,C^r_{31}+C^r_{6}\,\left(64\,a^2+32\right)+C^r_{11}\,\left(64\,a^2+32\right)\nonumber\\
&-C^r_{15}\,\left(64\,a^2+96\right)-C^r_{13}\,\left(64\,a^2+160\right)+C^r_{20}\,\left(64\,a^2+160\right)+C^r_{32}\,\left(64\,a^2+160\right)\notag\\
&+C^r_{21}\,\left(384\,a^2+192\right).\label{equ:A27}
\end{align}
and $d_{12,13}=m_\pi^6/F_\pi^6$. The scattering lengths $a_0^0$ and $a_0^2$ are related to $m_\pi^6\widetilde{C}_{8,9}/(32\pi F_\pi^6)$, respectively, where
\begin{align}
\widetilde{C}_8=&3A_{12}+2A_{13},\label{equ:A24}\\
\widetilde{C}_9=&A_{13}.\label{equ:A25}
\end{align}

For $\pi K$ scattering, the NNLO contribution is related to $T^{\frac{3}{2}}(s,t,u)$ and $T^{\frac{3}{2}}(u,t,s)$, where $s=(m_K+m_\pi)^2$, $t=0$, and $u=(m_K-m_\pi)^2$. They can  be written as Eq. \eqref{equ:A1} \cite{Bijnens:2004bu},
\begin{align}
\nonumber
A_{14}=&64\,C^r_{29}\,a^3-128\,C^r_{4}\,a^3-64\,C^r_{25}\,a^3-64\,C^r_{3}\,a^3-32\,C^r_{14}\,a^2\,\left(a+1\right)\\\nonumber
&-32\,C^r_{17}\,a\,\left(a^3+1\right)-16\,C^r_{15}\,a\,\left(4\,a^3+2\,a^2+3\,a+1\right)+16\,C^r_{1}\,a^2\,{\left(a+1\right)}^2\\\nonumber
&+64\,C^r_{2}\,a^2\,\left(a^2+1\right)+16\,C^r_{5}\,a^2\,\left(a^2+1\right)+32\,C^r_{7}\,a^2\,\left(a^2+1\right)+16\,C^r_{8}\,a^2\,\left(a^2+1\right)\\\nonumber
&+16\,C^r_{10}\,a^2\,\left(a^2+1\right)-48\,C^r_{12}\,a^2\,\left(a^2+1\right)-64\,C^r_{16}\,a^2\,\left(a^2+1\right)\\\nonumber
&+48\,C^r_{19}\,a^2\,\left(a^2+1\right)+32\,C^r_{22}\,a^2\,{\left(a+1\right)}^2+64\,C^r_{23}\,a^2\,\left(a^2+1\right)-16\,C^r_{26}\,a^2\,{\left(a+1\right)}^2\\\nonumber
&-64\,C^r_{28}\,a^2\,\left(a^2+1\right)+48\,C^r_{31}\,a^2\,\left(a^2+1\right)+32\,C^r_{6}\,a^2\,\left(2\,a^2+1\right)\\\nonumber
&+32\,C^r_{11}\,a^2\,\left(2\,a^2+1\right)-32\,C^r_{13}\,a^2\,\left(4\,a^2+3\right)+192\,C^r_{21}\,a^2\,\left(2\,a^2+1\right)\\
&+32\,C^r_{20}\,a^2\,\left(4\,a^2+3\right)+32\,C^r_{32}\,a^2\,\left(4\,a^2+3\right),\label{equ:A31}\\\nonumber
A_{15}=&64\,C^r_{3}\,a^3+128\,C^r_{4}\,a^3+64\,C^r_{25}\,a^3-64\,C^r_{29}\,a^3+32\,C^r_{14}\,a^2\,\left(a-1\right)\\\nonumber
&-32\,C^r_{17}\,a\,\left(a^3-1\right)-16\,C^r_{15}\,a\,\left(4\,a^3-2\,a^2+3\,a-1\right)+16\,C^r_{1}\,a^2\,{\left(a-1\right)}^2\\\nonumber
&+64\,C^r_{2}\,a^2\,\left(a^2+1\right)+16\,C^r_{5}\,a^2\,\left(a^2+1\right)+32\,C^r_{7}\,a^2\,\left(a^2+1\right)+16\,C^r_{8}\,a^2\,\left(a^2+1\right)\\\nonumber
&+16\,C^r_{10}\,a^2\,\left(a^2+1\right)-48\,C^r_{12}\,a^2\,\left(a^2+1\right)-64\,C^r_{16}\,a^2\,\left(a^2+1\right)\\\nonumber
&+48\,C^r_{19}\,a^2\,\left(a^2+1\right)+32\,C^r_{22}\,a^2\,{\left(a-1\right)}^2+64\,C^r_{23}\,a^2\,\left(a^2+1\right)-16\,C^r_{26}\,a^2\,{\left(a-1\right)}^2\\\nonumber
&-64\,C^r_{28}\,a^2\,\left(a^2+1\right)+48\,C^r_{31}\,a^2\,\left(a^2+1\right)+32\,C^r_{6}\,a^2\,\left(2\,a^2+1\right)\\\nonumber
&+32\,C^r_{11}\,a^2\,\left(2\,a^2+1\right)-32\,C^r_{13}\,a^2\,\left(4\,a^2+3\right)+192\,C^r_{21}\,a^2\,\left(2\,a^2+1\right)\\
&+32\,C^r_{20}\,a^2\,\left(4\,a^2+3\right)+32\,C^r_{32}\,a^2\,\left(4\,a^2+3\right),\label{equ:A32}
\end{align}
where $d_{14,15}=m_\pi^6/F_\pi^6$. The scattering lengths $a_0^{1/2}$ and $a_0^{3/2}$ are related to $m_\pi^6\widetilde{C}_{10,11}/(8 \pi F_\pi^6\sqrt{s})$, respectively, where
\begin{align}
\widetilde{C}_{10}=&-\frac{1}{2}A_{14}+\frac{3}{2}A_{15},\label{equ:A29}\\
\widetilde{C}_{11}=&A_{14}.\label{equ:A30}
\end{align}

For the pion scalar form factor $F_S^\pi(t)$, the NNLO contribution is \cite{Bijnens:2003xg}
\begin{align}
\Big(\frac{F_S^\pi(t)}{2B_0}\Big)_6=&\Big(\frac{F_S^\pi(t)}{2B_0}\Big)_{6,L}+\frac{1}{F_\pi^4}(A_{16}t^2+A_{17}tm_\pi^2+A_{18}m_\pi^4),\label{equ:A33}\\
A_{16}=&-8\,C^r_{12}-16\,C^r_{13},\label{equ:A34}\\
A_{17}=&32\,C^r_{12}+64\,C^r_{13}+16\,C^r_{14}+32\,C^r_{16}+16\,C^r_{17}+16\,C^r_{34}+16\,C^r_{36}+C^r_{15}\,\left(16\,a^2+24\right),\label{equ:A35}\\\nonumber
A_{18}=&144\,C^r_{19}-48\,C^r_{14}-48\,C^r_{17}-96\,C^r_{12}+96\,C^r_{31}-C^r_{15}\,\left(64\,a^2+64\right)\\\nonumber
&-C^r_{13}\,\left(128\,a^2+128\right)+C^r_{32}\,\left(128\,a^2+128\right)-C^r_{16}\,\left(64\,a^4-64\,a^2+112\right)\\
&+C^r_{20}\,\left(64\,a^4+64\,a^2+240\right)+C^r_{21}\,\left(192\,a^4+576\,a^2+240\right).\label{equ:A36}
\end{align}
$\langle r^2\rangle_S^\pi$ and $c_S^\pi$ are related to $\widetilde{C}_{12}=A_{16}$ and $\widetilde{C}_{13}=A_{17}$, respectively.

Reference \cite{Gasser:2007sg} gives the relations between $l_i^r$ and $L_i^r$ up to the NNLO. The NNLO contributions related to $C_i^r$ are $l_i^r\sim M_K^2\widetilde{C}_{i+13}/(16\pi^2 F_0^2),(i=1,2,3,4)$, where
\begin{align}
\widetilde{C}_{14}=&8C_6^r-8C_{11}^r+32C_{13}^r,\label{equ:A40}\\
\widetilde{C}_{15}=&16C_{11}^r-32C_{13}^r,\label{equ:A41}\\
\widetilde{C}_{16}=&-32C_{13}^r-16C_{15}^r+32C_{20}^r+192C_{21}^r+32C_{32}^r,\label{equ:A42}\\
\widetilde{C}_{17}^r=&16C_{15}^r,\label{equ:A43}
\end{align}
and $M_K^2$ is the one-loop expression of the kaon mass in the limit $m_u=m_d=0$ \cite{Gasser:1984gg}.

Now the number of $\widetilde{C}_i$ is related to the number of observables. They can be obtained directly.

\section{The values of $C_i^r$ in the other references}
\label{app:C}
\begin{longtable}{rrrrrrrr}
	\caption{$C_i^r$ in the other references. Some results with an asterisk mean the original results are $C_i$. We have reduced them to the renormalized ones. The numerical values are in units of $10^{-6}$.}\label{Ci}\\
	\hline\hline
	$i$&\multicolumn{7}{c}{$C_i^r$}\\
	\hline
	\endfirsthead
	
	\hline\hline
	$i$&\multicolumn{7}{c}{$C_i^r$}\\
	\hline
	\endhead
	
	\hline\hline
	\endfoot
	
	\hline\endlastfoot
	1&4.25$^*$\cite{Colangelo:2012ipa}&$-$2.55$^*$\cite{Colangelo:2012ipa}&$-$7.65$^*$\cite{Colangelo:2012ipa}&$-$16.15$^*$\cite{Colangelo:2012ipa}&$32.22^{+0.85}_{-1.45}$$^*$\cite{Jiang:2009uf}&30.69$^*$\cite{Jiang:2009uf}&12\cite{Bijnens:2014lea}\\
	1&$25.33^{+0.60}_{-1.11}$$^*$\cite{Jiang:2015dba}&12.16\cite{Bijnens:2011tb}&8.66\cite{Bijnens:2011tb}&16.83\cite{Bijnens:2011tb}&$-$7.33\cite{Bijnens:2011tb}&&\\
	2&$-$7.82$\pm$4.17$^*$\cite{Kampf:2006bn}&$-$6.29$\pm$4.17$^*$\cite{Kampf:2006bn}&$-$0.43$^*$\cite{Kampf:2006bn}&$0.00^{+0.00}_{-0.00}$$^*$\cite{Jiang:2009uf}&$\equiv$0$^*$\cite{Jiang:2009uf}&3.0\cite{Bijnens:2014lea}&$\equiv$0$^*$\cite{Jiang:2015dba}\\
	2&0.00\cite{Bijnens:2011tb}&1.13\cite{Bijnens:2011tb}&2.80\cite{Bijnens:2011tb}&&&&\\
	3&0.85$^*$\cite{Colangelo:2012ipa}&2.55$^*$\cite{Colangelo:2012ipa}&3.40$^*$\cite{Colangelo:2012ipa}&5.95$^*$\cite{Colangelo:2012ipa}&$-0.43^{+0.09}_{-0.09}$$^*$\cite{Jiang:2009uf}&$-$0.09$^*$\cite{Jiang:2009uf}&4.0\cite{Bijnens:2014lea}\\
	3&$-0.43^{+0.09}_{-0.09}$$^*$\cite{Jiang:2015dba}&0.00\cite{Bijnens:2011tb}&$-$0.11\cite{Bijnens:2011tb}&3.24\cite{Bijnens:2011tb}&0.84\cite{Bijnens:2011tb}&&\\
	4&5.10$^*$\cite{Colangelo:2012ipa}&0$^*$\cite{Colangelo:2012ipa}&$-$4.25$^*$\cite{Colangelo:2012ipa}&$-$10.20$^*$\cite{Colangelo:2012ipa}&$26.35^{+0.77}_{-1.28}$$^*$\cite{Jiang:2009uf}&25.33$^*$\cite{Jiang:2009uf}&15\cite{Bijnens:2014lea}\\
	4&$18.11^{+0.51}_{-0.85}$$^*$\cite{Jiang:2015dba}&14.52\cite{Bijnens:2011tb}&7.08\cite{Bijnens:2011tb}&22.25\cite{Bijnens:2011tb}&12.66\cite{Bijnens:2011tb}&&\\
	5&$-8.59^{+0.68}_{-0.94}$$^*$\cite{Jiang:2009uf}&$-$4.34$^*$\cite{Jiang:2009uf}&$-$4.0\cite{Bijnens:2014lea}&$-10.88^{+0.85}_{-1.11}$$^*$\cite{Jiang:2015dba}&6.19\cite{Bijnens:2011tb}&$-$2.31\cite{Bijnens:2011tb}&7.79\cite{Bijnens:2011tb}\\
	5&11.47\cite{Bijnens:2011tb}&&&&&&\\
	6&$\equiv$0$^*$\cite{Jiang:2009uf}&$-$4.0\cite{Bijnens:2014lea}&$\equiv$0$^*$\cite{Jiang:2015dba}&0.00\cite{Bijnens:2011tb}&$-$3.07\cite{Bijnens:2011tb}&$-$0.50\cite{Bijnens:2011tb}&\\
	7&$\equiv$0$^*$\cite{Jiang:2009uf}&5.0\cite{Bijnens:2014lea}&$\equiv$0$^*$\cite{Jiang:2015dba}&0.00\cite{Bijnens:2011tb}&3.50\cite{Bijnens:2011tb}&$-$0.03\cite{Bijnens:2011tb}&\\
	8&$19.64^{-1.36}_{+1.53}$$^*$\cite{Jiang:2009uf}&9.86$^*$\cite{Jiang:2009uf}&19\cite{Bijnens:2014lea}&$17.85^{-1.28}_{+1.36}$$^*$\cite{Jiang:2015dba}&6.19\cite{Bijnens:2011tb}&5.28\cite{Bijnens:2011tb}&14.34\cite{Bijnens:2011tb}\\
	8&6.15\cite{Bijnens:2011tb}&&&&&&\\
	10&$-8.93^{+0.68}_{-0.77}$$^*$\cite{Jiang:2009uf}&$-$4.17$^*$\cite{Jiang:2009uf}&$-$0.25\cite{Bijnens:2014lea}&$-5.53^{+0.43}_{-0.51}$$^*$\cite{Jiang:2015dba}&$-$12.39\cite{Bijnens:2011tb}&$-$2.40\cite{Bijnens:2011tb}&$-$1.64\cite{Bijnens:2011tb}\\
	10&3.07\cite{Bijnens:2011tb}&&&&&&\\
	11&$\equiv$0$^*$\cite{Jiang:2009uf}&$-$4.0\cite{Bijnens:2014lea}&$\equiv$0$^*$\cite{Jiang:2015dba}&0.00\cite{Bijnens:2011tb}&$-$1.12\cite{Bijnens:2011tb}&$-$3.44\cite{Bijnens:2011tb}&\\
	12&0.03$\pm$0.54\cite{Jamin:2004re}&$-$10\cite{Bijnens:2003uy}&$-$3.74$\pm$1.36$^*$\cite{Cirigliano:2005xn}&$-$0.6$\pm$0.3\cite{Unterdorfer:2008zz}&4.90$\pm$0.48\cite{Bernard:2009ds}&6.66$\pm$0.49\cite{Bernard:2009ds}&3.99$\pm$0.81\cite{Bernard:2009ds}\\
	12&4.88$\pm$0.81\cite{Bernard:2009ds}&3.99$\pm$0.48\cite{Bernard:2009ds}&3.77$\pm$0.75\cite{Bernard:2009ds}&0.43$\pm$0.34$^*$\cite{Golterman:2014nua}&$-2.89^{+0.17}_{-0.09}$$^*$\cite{Jiang:2009uf}&$-$1.62$^*$\cite{Jiang:2009uf}&$-$2.8\cite{Bijnens:2014lea}\\
	12&$-$2.4\cite{Bijnens:2014lea,Cirigliano:2006hb}&$-$0.421\cite{Bernard:2007tk}&$-$0.484\cite{Bernard:2007tk}&$-$0.550\cite{Bernard:2007tk}&$-$0.362\cite{Bernard:2007tk}&$-$0.306\cite{Bernard:2007tk}&$-$0.170\cite{Bernard:2007tk}\\
	12&$-$0.235\cite{Bernard:2007tk}&$-$0.683\cite{Bernard:2007tk}&$-$0.743\cite{Bernard:2007tk}&$-$0.234\cite{Bernard:2007tk}&1.107\cite{Bernard:2007tk}&$-$0.202\cite{Bernard:2007tk}&1.132\cite{Bernard:2007tk}\\
	12&$-$0.264\cite{Bernard:2007tk}&1.084\cite{Bernard:2007tk}&$-$15\cite{Bijnens:2003xg}&$-$5.2\cite{Bijnens:2003xg}&2.6\cite{Bijnens:2003xg}&7.8\cite{Bijnens:2003xg}&$-$11\cite{Bijnens:2003xg}\\
	12&$-$8.4\cite{Bijnens:2003xg}&1.2\cite{Bijnens:2003xg}&$-$13\cite{Bijnens:2003xg}&$-2.89^{+0.09}_{-0.09}$$^*$\cite{Jiang:2015dba}&$-$6.19\cite{Bijnens:2011tb}&$-$0.78\cite{Bijnens:2011tb}&$-$13.58\cite{Bijnens:2011tb}\\
	12&$-$5.12\cite{Bijnens:2011tb}&&&&&&\\
	13&0$\pm$0.2\cite{Unterdorfer:2008zz}&$\equiv$0$^*$\cite{Jiang:2009uf}&1.5\cite{Bijnens:2014lea}&$-$5.6\cite{Bijnens:2003xg}&$-$0.2\cite{Bijnens:2003xg}&1.5\cite{Bijnens:2003xg}&0.3\cite{Bijnens:2003xg}\\
	13&$\equiv$0$^*$\cite{Jiang:2015dba}&0.00\cite{Bijnens:2011tb}&2.65\cite{Bijnens:2011tb}&$-$0.02\cite{Bijnens:2011tb}&&&\\
	14&$-$36.55$^*$\cite{Amoros:1999dp}&$\equiv$0\cite{Bernard:2009ds}&0.60$\pm$1.21\cite{Bernard:2009ds}&0.55$\pm$1.17\cite{Bernard:2009ds}&$\equiv$0.55\cite{Bernard:2009ds}&$-$0.79$\pm$0.57\cite{Bernard:2009ds}&$-7.06^{+1.02}_{-1.62}$$^*$\cite{Jiang:2009uf}\\
	14&$-$2.21$^*$\cite{Jiang:2009uf}&$-$1.0\cite{Bijnens:2014lea}&$-7.40^{+1.19}_{-1.79}$$^*$\cite{Jiang:2015dba}&0.00\cite{Bijnens:2011tb}&$-$1.90\cite{Bijnens:2011tb}&$-$7.59\cite{Bijnens:2011tb}&$-$8.28\cite{Bijnens:2011tb}\\
	15&$\equiv$0$^*$\cite{Jiang:2009uf}&$-$3.0\cite{Bijnens:2014lea}&$\equiv$0$^*$\cite{Jiang:2015dba}&0.00\cite{Bijnens:2011tb}&$-$2.28\cite{Bijnens:2011tb}&$-$2.33\cite{Bijnens:2011tb}&\\
	16&$\equiv$0$^*$\cite{Jiang:2009uf}&3.2\cite{Bijnens:2014lea}&2.4\cite{Moussallam:2000zf}&2.6\cite{Moussallam:2000zf}&2.8\cite{Moussallam:2000zf}&3.2\cite{Moussallam:2000zf}&$\equiv$0$^*$\cite{Jiang:2015dba}\\
	16&0.00\cite{Bijnens:2011tb}&0.07\cite{Bijnens:2011tb}&0.63\cite{Bijnens:2011tb}&&&&\\
	17&$\equiv$0\cite{Bernard:2009ds}&0.13$\pm$1.41\cite{Bernard:2009ds}&0.82$\pm$1.43\cite{Bernard:2009ds}&$\equiv$0.13\cite{Bernard:2009ds}&1.77$\pm$0.66\cite{Bernard:2009ds}&$0.09^{-0.09}_{-0.09}$$^*$\cite{Jiang:2009uf}&$-$1.28$^*$\cite{Jiang:2009uf}\\
	17&$-$1.0\cite{Bijnens:2014lea}&$1.45^{+0.09}_{-0.34}$$^*$\cite{Jiang:2015dba}&0.00\cite{Bijnens:2011tb}&0.02\cite{Bijnens:2011tb}&1.25\cite{Bijnens:2011tb}&11.21\cite{Bijnens:2011tb}&\\
	18&$-4.76^{+0.77}_{-0.94}$$^*$\cite{Jiang:2009uf}&$-$0.51$^*$\cite{Jiang:2009uf}&0.63\cite{Bijnens:2014lea}&$-$1.8\cite{Bijnens:2014lea,Kaiser:2007zz}&$-5.10^{+0.60}_{-0.77}$$^*$\cite{Jiang:2015dba}&$-$2.02\cite{Bijnens:2011tb}&$-$1.28\cite{Bijnens:2011tb}\\
	18&$-$2.84\cite{Bijnens:2011tb}&$-$0.63\cite{Bijnens:2011tb}&&&&&\\
	19&$-$23.80$^*$\cite{Amoros:1999dp}&$-4.08^{+0.77}_{-1.11}$$^*$\cite{Jiang:2009uf}&$-$0.68$^*$\cite{Jiang:2009uf}&$-$4.0\cite{Bijnens:2014lea}&$-$0.6\cite{Bijnens:2014lea,Kaiser:2007zz}&$-$1.7\cite{Moussallam:2000zf}&$-$3.4\cite{Moussallam:2000zf}\\
	19&$-$4.5\cite{Moussallam:2000zf}&$-$3.8\cite{Moussallam:2000zf}&$-$2.4\cite{Moussallam:2000zf}&$-2.30^{+0.77}_{-1.11}$$^*$\cite{Jiang:2015dba}&0.01\cite{Bijnens:2011tb}&$-$1.10\cite{Bijnens:2011tb}&$-$4.11\cite{Bijnens:2011tb}\\
	19&$-$11.47\cite{Bijnens:2011tb}&&&&&&\\
	20&$1.53^{-0.26}_{+0.34}$$^*$\cite{Jiang:2009uf}&0.17$^*$\cite{Jiang:2009uf}&1.0\cite{Bijnens:2014lea}&0.9\cite{Bijnens:2014lea,Kaiser:2007zz}&$-$0.5\cite{Moussallam:2000zf}&0.7\cite{Moussallam:2000zf}&1.2\cite{Moussallam:2000zf}\\
	20&0.8\cite{Moussallam:2000zf}&0.4\cite{Moussallam:2000zf}&$1.45^{-0.17}_{+0.26}$$^*$\cite{Jiang:2015dba}&$-$0.02\cite{Bijnens:2011tb}&0.41\cite{Bijnens:2011tb}&$-$3.35\cite{Bijnens:2011tb}&$-$0.43\cite{Bijnens:2011tb}\\
	21&$-0.51^{+0.09}_{-0.09}$$^*$\cite{Jiang:2009uf}&$-$0.09$^*$\cite{Jiang:2009uf}&$-$0.48\cite{Bijnens:2014lea}&$-0.51^{+0.09}_{-0.09}$$^*$\cite{Jiang:2015dba}&0.01\cite{Bijnens:2011tb}&$-$0.14\cite{Bijnens:2011tb}&0.18\cite{Bijnens:2011tb}\\
	21&$-$0.88\cite{Bijnens:2011tb}&&&&&&\\
	22&$2.30^{+1.62}_{-2.13}$$^*$\cite{Jiang:2009uf}&9.44$^*$\cite{Jiang:2009uf}&9.0\cite{Bijnens:2014lea}&$-2.98^{+1.70}_{-2.21}$$^*$\cite{Jiang:2015dba}&$-$2.97\cite{Bijnens:2011tb}&0.62\cite{Bijnens:2011tb}&5.45\cite{Bijnens:2011tb}\\
	22&11.17\cite{Bijnens:2011tb}&&&&&&\\
	23&$\equiv$0$^*$\cite{Jiang:2009uf}&$-$1.0\cite{Bijnens:2014lea}&$\equiv$0$^*$\cite{Jiang:2015dba}&0.00\cite{Bijnens:2011tb}&0.48\cite{Bijnens:2011tb}&2.69\cite{Bijnens:2011tb}&\\
	25&$-50.84^{-4.17}_{+6.12}$$^*$\cite{Jiang:2009uf}&$-$61.29$^*$\cite{Jiang:2009uf}&$-$11\cite{Bijnens:2014lea}&$-25.76^{-3.49}_{+5.02}$$^*$\cite{Jiang:2015dba}&$-$18.38\cite{Bijnens:2011tb}&$-$13.66\cite{Bijnens:2011tb}&$-$14.52\cite{Bijnens:2011tb}\\
	25&12.82\cite{Bijnens:2011tb}&&&&&&\\
	26&$28.48^{+2.47}_{-4.00}$$^*$\cite{Jiang:2009uf}&33.41$^*$\cite{Jiang:2009uf}&10\cite{Bijnens:2014lea}&$23.04^{+2.98}_{-4.59}$$^*$\cite{Jiang:2015dba}&$-$2.84\cite{Bijnens:2011tb}&7.65\cite{Bijnens:2011tb}&$-$5.97\cite{Bijnens:2011tb}\\
	26&$-$4.85\cite{Bijnens:2011tb}&&&&&&\\
	28&$2.55^{+0.09}_{-0.09}$$^*$\cite{Jiang:2009uf}&2.47$^*$\cite{Jiang:2009uf}&$-$2.0\cite{Bijnens:2014lea}&$1.53^{+0.00}_{-0.09}$$^*$\cite{Jiang:2015dba}&1.35\cite{Bijnens:2011tb}&0.69\cite{Bijnens:2011tb}&1.77\cite{Bijnens:2011tb}\\
	28&1.47\cite{Bijnens:2011tb}&&&&&&\\
	29&$-26.18^{-2.21}_{+2.72}$$^*$\cite{Jiang:2009uf}&$-$32.39$^*$\cite{Jiang:2009uf}&$-$20\cite{Bijnens:2014lea}&$-8.42^{-1.79}_{+2.04}$$^*$\cite{Jiang:2015dba}&$-$13.63\cite{Bijnens:2011tb}&$-$7.04\cite{Bijnens:2011tb}&$-$19.07\cite{Bijnens:2011tb}\\
	29&$-$7.85\cite{Bijnens:2011tb}&&&&&&\\
	30&$5.10^{+0.17}_{-0.26}$$^*$\cite{Jiang:2009uf}&4.93$^*$\cite{Jiang:2009uf}&3.0\cite{Bijnens:2014lea}&$3.15^{+0.09}_{-0.17}$$^*$\cite{Jiang:2015dba}&2.70\cite{Bijnens:2011tb}&1.37\cite{Bijnens:2011tb}&1.65\cite{Bijnens:2011tb}\\
	30&5.45\cite{Bijnens:2011tb}&&&&&&\\
	31&$-5.36^{+0.43}_{-0.77}$$^*$\cite{Jiang:2009uf}&$-$1.87$^*$\cite{Jiang:2009uf}&2.0\cite{Bijnens:2014lea}&$-3.91^{+0.60}_{-1.11}$$^*$\cite{Jiang:2015dba}&$-$6.16\cite{Bijnens:2011tb}&$-$1.44\cite{Bijnens:2011tb}&$-$3.89\cite{Bijnens:2011tb}\\
	31&13.10\cite{Bijnens:2011tb}&&&&&&\\
	32&$1.53^{-0.26}_{+0.34}$$^*$\cite{Jiang:2009uf}&0.17$^*$\cite{Jiang:2009uf}&1.7\cite{Bijnens:2014lea}&$1.45^{-0.17}_{+0.26}$$^*$\cite{Jiang:2015dba}&$-$0.02\cite{Bijnens:2011tb}&0.41\cite{Bijnens:2011tb}&2.91\cite{Bijnens:2011tb}\\
	32&3.56\cite{Bijnens:2011tb}&&&&&&\\
	33&$0.77^{-0.00}_{+0.26}$$^*$\cite{Jiang:2009uf}&0.68$^*$\cite{Jiang:2009uf}&0.82\cite{Bijnens:2014lea}&$-0.43^{-0.17}_{+0.43}$$^*$\cite{Jiang:2015dba}&2.08\cite{Bijnens:2011tb}&0.21\cite{Bijnens:2011tb}&2.91\cite{Bijnens:2011tb}\\
	33&$-$1.02\cite{Bijnens:2011tb}&&&&&&\\
	34&5.61$\pm$4.00$^*$\cite{Cirigliano:2005xn}&2.16$\pm$0.37\cite{Bernard:2009ds}&$-$1.09$\pm$0.37\cite{Bernard:2009ds}&3.20$\pm$0.81\cite{Bernard:2009ds}&0.91$\pm$0.82\cite{Bernard:2009ds}&3.20$\pm$0.37\cite{Bernard:2009ds}&2.98$\pm$0.80\cite{Bernard:2009ds}\\
	34&$13.52^{-0.85}_{+1.36}$$^*$\cite{Jiang:2009uf}&8.76$^*$\cite{Jiang:2009uf}&7.0\cite{Bijnens:2014lea}&6.480\cite{Bernard:2007tk}&3.971\cite{Bernard:2007tk}&1.344\cite{Bernard:2007tk}&8.879\cite{Bernard:2007tk}\\
	34&11.176\cite{Bernard:2007tk}&4.741\cite{Bernard:2007tk}&2.235\cite{Bernard:2007tk}&8.229\cite{Bernard:2007tk}&5.718\cite{Bernard:2007tk}&1.534\cite{Bernard:2007tk}&$-$0.216\cite{Bernard:2007tk}\\
	34&0.666\cite{Bernard:2007tk}&$-$1.092\cite{Bernard:2007tk}&2.400\cite{Bernard:2007tk}&0.659\cite{Bernard:2007tk}&$5.61^{-1.53}_{+2.47}$$^*$\cite{Jiang:2015dba}&14.32\cite{Bijnens:2011tb}&3.63\cite{Bijnens:2011tb}\\
	34&23.21\cite{Bijnens:2011tb}&10.77\cite{Bijnens:2011tb}&&&&&\\
	36&$\equiv$0$^*$\cite{Jiang:2009uf}&2.0\cite{Bijnens:2014lea}&$\equiv$0$^*$\cite{Jiang:2015dba}&0.00\cite{Bijnens:2011tb}&3.89\cite{Bijnens:2011tb}&$-$0.95\cite{Bijnens:2011tb}&\\
	63&$25.42^{-2.04}_{+2.55}$$^*$\cite{Jiang:2009uf}&11.98$^*$\cite{Jiang:2009uf}&$21.08^{-1.79}_{+2.13}$$^*$\cite{Jiang:2015dba}&6.19\cite{Bijnens:2011tb}&6.83\cite{Bijnens:2011tb}&6.65\cite{Bijnens:2011tb}&7.76\cite{Bijnens:2011tb}\\
	66&3.40$^*$\cite{Colangelo:2012ipa}&$-$2.55$^*$\cite{Colangelo:2012ipa}&$-$5.95$^*$\cite{Colangelo:2012ipa}&$-$12.75$^*$\cite{Colangelo:2012ipa}&$14.54^{+0.60}_{-1.02}$$^*$\cite{Jiang:2009uf}&14.71$^*$\cite{Jiang:2009uf}&$0.68^{+0.34}_{-0.60}$$^*$\cite{Jiang:2015dba}\\
	66&10.49\cite{Bijnens:2011tb}&3.91\cite{Bijnens:2011tb}&17.03\cite{Bijnens:2011tb}&4.16\cite{Bijnens:2011tb}&&&\\
	69&$-$3.40$^*$\cite{Colangelo:2012ipa}&2.55$^*$\cite{Colangelo:2012ipa}&5.95$^*$\cite{Colangelo:2012ipa}&12.75$^*$\cite{Colangelo:2012ipa}&$-7.31^{-0.34}_{+0.51}$$^*$\cite{Jiang:2009uf}&$-$7.65$^*$\cite{Jiang:2009uf}&$4.42^{+0.00}_{-0.09}$$^*$\cite{Jiang:2015dba}\\
	69&$-$5.77\cite{Bijnens:2011tb}&$-$1.96\cite{Bijnens:2011tb}&$-$6.64\cite{Bijnens:2011tb}&$-$7.84\cite{Bijnens:2011tb}&&&\\
	83&$0.60^{+1.70}_{-2.30}$$^*$\cite{Jiang:2009uf}&8.16$^*$\cite{Jiang:2009uf}&$-14.79^{+1.45}_{-1.87}$$^*$\cite{Jiang:2015dba}&1.63\cite{Bijnens:2011tb}&0.16\cite{Bijnens:2011tb}&$-$2.94\cite{Bijnens:2011tb}&$-$5.53\cite{Bijnens:2011tb}\\
	88&$-$52$^*$\cite{Knecht:2001xc}&$-$16$^*$\cite{Knecht:2001xc}&$-$14$^*$\cite{Knecht:2001xc}&$-$3.5$\pm$1.0\cite{Unterdorfer:2008zz}&$-46.50^{-6.21}_{+8.76}$$^*$\cite{Jiang:2009uf}&$-$66.56$^*$\cite{Jiang:2009uf}&$-14.37^{-5.78}_{+7.91}$$^*$\cite{Jiang:2015dba}\\
	88&$-$13.83\cite{Bijnens:2011tb}&$-$12.49\cite{Bijnens:2011tb}&$-$9.12\cite{Bijnens:2011tb}&$-$3.31\cite{Bijnens:2011tb}&&&\\
	90&0.0$^*$\cite{Knecht:2001xc}&33$^*$\cite{Knecht:2001xc}&51$^*$\cite{Knecht:2001xc}&$20.74^{-3.23}_{+3.91}$$^*$\cite{Jiang:2009uf}&2.13$^*$\cite{Jiang:2009uf}&$19.72^{-3.74}_{+4.68}$$^*$\cite{Jiang:2015dba}&50.69\cite{Bijnens:2011tb}\\
	90&5.57\cite{Bijnens:2011tb}&52.38\cite{Bijnens:2011tb}&$-$2.04\cite{Bijnens:2011tb}&&&&\\
	\hline\hline
\end{longtable}

\begin{table}[h]
	\begin{ruledtabular}
  \caption{The initial intervals of $C_{i}^r$. These values are calculated by Eq. \eqref{ic} and some outliers are excluded. }\label{cb}
		\begin{tabular}{cccccccc}
			$i$ &     $C_i^r$      & $i$ &     $C_i^r$      & $i$ &     $C_i^r$      & $i$ &    $C_i^r$     \\
			\hline
			 1  &   9.0$\pm$15.6   & 12  &  $-$1.6$\pm$5.1  & 22  &   4.0$\pm$5.6    & 34  &  4.7$\pm$4.2   \\
			 2  & $-$0.76$\pm$3.55 & 13  & 0.012$\pm$2.074  & 23  &  0.36$\pm$1.24   & 36  & 0.82$\pm$1.79  \\
			 3  &   1.6$\pm$2.1    & 14  &  $-$2.7$\pm$3.5  & 25  &   $-$23$\pm$23   & 63  &    12$\pm$8    \\
			 4  &    11$\pm$12     & 15  &  $-$1.3$\pm$1.4  & 26  &    11$\pm$16     & 66  &  4.3$\pm$9.3   \\
			 5  & $-$0.58$\pm$8.11 & 16  &   1.5$\pm$1.5    & 28  &   1.2$\pm$1.4    & 69  & $-$1.4$\pm$6.9 \\
			 6  &  $-$1.3$\pm$1.8  & 17  &  0.28$\pm$0.91   & 29  &   $-$17$\pm$9    & 83  & $-$1.8$\pm$7.1 \\
			 7  &   1.4$\pm$2.2    & 18  &  $-$2.0$\pm$1.9  & 30  &   3.4$\pm$1.6    & 88  &  $-$23$\pm$22  \\
			 8  &     12$\pm$6     & 19  &  $-$3.2$\pm$2.8  & 31  & $-$0.94$\pm$6.22 & 90  &   23$\pm$22    \\
			10  &  $-$4.0$\pm$4.9  & 20  &  0.30$\pm$1.23   & 32  &   1.5$\pm$1.3    &     &                \\
			11  &  $-$1.4$\pm$1.8  & 21  & $-$0.30$\pm$0.35 & 33  &  0.75$\pm$1.27   &     &
		\end{tabular}
	\end{ruledtabular}
\end{table}

\end{document}